\documentclass[letterpaper, 11pt]{amsart} 

\usepackage{fullpage}
\usepackage{lipsum} 
\usepackage{siunitx}
\usepackage{amsfonts}
\usepackage{graphicx} 
\usepackage{epstopdf} 
\usepackage{mathtools} 
\usepackage{bm} 
\usepackage[caption=false]{subfig} 
\usepackage{tikz} 
\usepackage{stmaryrd}
\usetikzlibrary{arrows.meta} 
\usepackage{color} 

\allowdisplaybreaks

\usepackage{url} 
\usepackage[plainpages = false, colorlinks, citecolor = blue, filecolor = black, linkcolor = red, urlcolor = blue]{hyperref} 
\usepackage[capitalize]{cleveref}
\crefname{subsection}{Subsection}{Subsections}
\crefname{equation}{Equation}{Equations}
\crefname{figure}{Figure}{Figures}

\newcommand{\ep}{\varepsilon}
\newcommand{\R}{\mathbb{R}} 
\newcommand{\Wo}{\textrm{Wo}}
  
\newcommand{\gp}{\nabla^{\perp}}
\newcommand{\ds}{\displaystyle} 
\newcommand{\x}{\mathbf{x}} 
\newcommand{\vect}[1]{\bm{#1}}
\renewcommand{\u}{\vect{u}}
\newcommand{\vor}{\vect{\omega}}
\newcommand{\unitvect}[1]{\hat{\vect{#1}}}
\renewcommand{\i}{\unitvect{\imath}}
\renewcommand{\j}{\unitvect{\jmath}}
\renewcommand{\k}{\unitvect{k}}
\renewcommand{\Re}[1]{\text{Real}\left\{#1\right\}}
\renewcommand{\Im}[1]{\text{Imag}\left\{#1\right\}}

\newcommand{\myblue}{blue!80!black}
\newcommand{\myred}{red!85!black}
\newcommand{\mygreen}{green!45!black}

\ifpdf
  \DeclareGraphicsExtensions{.eps,.pdf,.png,.jpg}
\else
  \DeclareGraphicsExtensions{.eps}
\fi

\theoremstyle{definition}
\newtheorem{definition}{Definition}[section]
\newtheorem{remark}[definition]{Remark}

\usepackage{amsopn}

\usepackage[backend = bibtex, giveninits = true, style = alphabetic, natbib = true, url = false, sorting = nyt, doi = true, abbreviate = true, backref = false]{biblatex}
\renewbibmacro{in:}{\ifentrytype{article}{}{\printtext{\bibstring{in}\intitlepunct}}}

\begin{document}
\title[3D Viscous Steady Streaming]{Three-dimensional viscous steady streaming in a rectangular channel past a cylinder}

\author{Nathan Willis}
\address{Department of Applied Mathematics, University of California, Merced, Merced, CA}
\email{nwillis@ucmerced.edu}
\author{Christel Hohenegger}
\address{Department of Mathematics, University of Utah, Salt Lake City, UT}
\email{choheneg@math.utah.edu} 


\date{\today}

\begin{abstract} 
We consider viscous steady streaming induced by oscillatory flow past a cylinder between two plates, where the cylinder's axis is normal to the plates. While this phenomenon was first studied in the 1930s, it has received renewed interest recently for possible applications in particle manipulations and non-Newtonian flows. The flow is driven at the ends of the channel by the  boundary condition which is a series solution of the oscillating flow problem in a rectangular channel in the absence of a cylinder. We use a combination of Fourier series and an asymptotic expansion to study the confinement effects for steady-streaming. The Fourier series in time naturally simplifies to a finite series. In contrast, it is necessary to truncate the Fourier series in $z$, which is in the direction of the axis of the cylinder, to solve numerically. The successive equations for the Fourier coefficients resulting from the asymptotic expansion are then solved numerically using finite element methods. We use our model to evaluate how steady streaming depends on the domain width and distance from the cylinder to the outer walls, including the possible breaking of the four-fold symmetry due to the domain shape. We utilize the tangential steady-streaming velocity along the radial chord at an angle of $\frac{\pi}{4}$ to analyze our solutions over an extensive range of oscillating frequencies and multiple levels in the $z$-direction. Finally, higher-order solutions are computed and an asymptotic correction to steady streaming is included.
\end{abstract}

\maketitle


\section{Introduction} \label{sec:intro}
Steady streaming refers to the time-independent component of the secondary flow superimposed on the primary oscillatory flow. While steady streaming can be generated in oscillatory flow past a variety of objects, we are specifically interested in steady streaming induced by oscillatory flow past a cylinder. Physically, we envision submerging tracer particles in the fluid, which is then oscillated at the ends of whatever apparatus contains the fluid. If one were to watch the tracer particles, they would see the primary flow, which is simply the fluid moving back and forth around the cylinder, rather uninterestingly. In contrast, steady streaming is observed by rapidly imaging either at the driving frequency or a much higher frequency and only considering the images of the tracer particles at the same frequency as the driving oscillation. In this way, the observer will only see the non-periodic component of the secondary flow. 

As is the case for a plethora of distinct phenomena in fluid dynamics, steady (or acoustic) streaming was first observed and investigated by Faraday \cite{faraday1831acoustic} and Rayleigh \cite{rayleigh1884circulation}. However, it was Schlichting \cite{schlichting1932berechnung} who first studied the problem of steady streaming induced by oscillatory flow past a cylinder. Serendipitously, Schlichting \cite{schlichting1932berechnung} was extending the classical boundary layer analysis of a viscous fluid by Prandtl \cite{prandtl1904friction} to oscillatory flow when it was determined that the solutions at the next order in the asymptotic expansion have a time-independent component. For the next six decades, boundary layers were used to further develop theory and analysis for steady streaming, see e.g. \cite{stuart1966double,riley1967oscillatory,wang1968high,kim1989streaming}. In 1973, Bertelsen, Svadal, and Tj{\o}tta \cite{bertelsen1973nonlinear} split the velocity into steady and unsteady parts to handle the boundary conditions on the cylinder to re-evaluate the valid regime of Reynolds numbers determined by Holtsmark et al. \cite{holtsmark1954boundary}. Further, Bertelsen et al. \cite{bertelsen1973nonlinear} solved for the steady streaming between two co-axial cylinders and compared their results to experiments. In this paper, we consider the case of small amplitude oscillations, as compared to cylinder radius, in a viscous fluid, while the instability or flow separation  occur for very large amplitude oscillations or very low viscosity fluids, see \cite{honji1981streaked,hall1984stability,tatsuno1990visual,sarpkaya2002experiments} for details. 

Another approach, which is more in line with the work presented here, does not explicitly split the inner and outer layers or split into steady and unsteady parts. Riley \cite{andres1953acoustic} used a perturbation from the oscillating flow to study steady streaming at Reynolds number around an order of ten. Holtsmark et al. \cite{holtsmark1954boundary} used successive approximations and special functions to solve for the steady streaming solution in an infinite planar domain and compared their results to their experimental results. Kubo and Kitano \cite{kubo1980secondary} similarly employed successive approximations to study steady streaming induced by a cylinder oscillating in two directions.

In 2005, Lutz, Chen, and Schwartz \cite{lutz2005microscopic} experimentally investigated steady streaming for applications in microfluidic devices. They imaged the flow in the plane which is orthogonal to the cylinder, which is standard for steady streaming experiments. They also imaged the flow in the planes that bisect the cylinder along the axis and are parallel to the channel walls. Recently, in 2019, Vishwanathan and Juarez \cite{vishwanathan2019steadyViscometry} utilized small-amplitude, high-frequency, steady streaming to experimentally measure viscosities of Newtonian fluids for applications in microfluidic devices. In the last two decades, steady streaming has been used in the chemical engineering and chemistry fields for particle manipulation in fluids without any physical contact with the particles. That is, without directly interacting with particles, they can mix \cite{lutz2006characterizing}, trap \cite{lieu2012hydrodynamic,parthasarathy2019streaming,volk2020size}, and sort \cite{thameem2017fast, wang2011size} a variety of microparticles in the fluid via steady streaming induced by a circular cylinder. For example, in 2012, a device was manufactured by Lieu, House, and Schwartz \cite{lieu2012hydrodynamic} such that the steady streaming flow would trap particles in specific locations. These experiments are completed on the micron scale, suggesting these methods for particle manipulation are directly applicable to the study of biological fluids. Theoretically, Chong et al. \cite{chong2013inertial,chong2016transport} have shown the applicability of steady streaming for particle trapping or transport via a single oscillating cylinder or an array of oscillating cylinders. For the purpose of particle trapping or transport, Bhosale et al. \cite{bhosale2020shape} conducted a bifurcation analysis on direct numerical simulations for two-dimensional steady streaming considering flow past non-circular cylinders and a regular lattice of circular cylinders with multiple length scales.


Considering three-dimensional steady streaming we note that there has been a significant amount of work focused on spheres or spheroids \cite{lane1955acoustical,wang1965flow,riley1966sphere,riley1967oscillatory,dohara1982unsteady,otto1992stability,hall1984stability,kong2017oscillatory,kotas2007visualization}. However, there is little theoretical work on three-dimensional flow past a cylinder. In 2015, Rallabandi et al. \cite{rallabandi2015three} considered three-dimensional steady streaming past a bubble, which allows slip, in a rectangular channel to investigate confinement effects. Very recently, Zhang and Rallabandi \cite{zhang2023three} considered a Hele-Shaw set up and used lubrication theory to study the impact of the channel having a vertical length scale much smaller than cylinder radius. Within this framework they separated the flow to an inner and outer flow, i.e., near and far from the cylinder, and primarily focus on the outer flow to show that the flow reverses direction throughout the height of the channel.

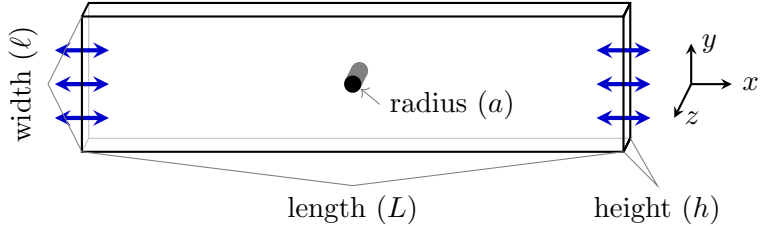
\begin{figure} 
\centering
\begin{tikzpicture}[scale = 0.9,  every node/.style={scale=1}];
\def\xend{8};
\def\yend{2};
\def\xshift{0.1};
\def\yshift{0.2};
\def\radd{0.5*sqrt{(\xshift*\xshift + \yshift*\yshift)}};
\def\cirxshift{0.894427190999915878563669467492510494176247343844610289708};
\def\ciryshift{0.447213595499957939281834733746255247088123671922305144854};

\draw[color = lightgray] (\xshift, \yshift) rectangle (\xend + \xshift, \yend + \yshift);
\draw[color = lightgray] (0, 0) -- (\xshift, \yshift);
\draw[ultra thick, color = \myblue, stealth-stealth] (-0.4,\yend/4) -- (0.4,\yend/4);
\draw[ultra thick, color = \myblue, stealth-stealth] (-0.4,\yend/2) -- (0.4,\yend/2);
\draw[ultra thick, color = \myblue, stealth-stealth] (-0.4,3*\yend/4) -- (0.4,3*\yend/4);

\draw[thick, color = black] (\xend,0) -- (\xend+ \xshift, 0 + \yshift) -- (\xend + \xshift, \yend + \yshift) -- (0 + \xshift, \yend + \yshift) -- (0,\yend) -- (0,0) -- (\xend,0) -- (\xend+\xshift, 0 + \yshift);

\draw[ultra thick, color = \myblue, stealth-stealth] (\xend -0.4,\yend/4) -- (\xend + 0.4,\yend/4);
\draw[ultra thick, color = \myblue, stealth-stealth] (\xend -0.4,\yend/2) -- (\xend + 0.4,\yend/2);
\draw[ultra thick, color = \myblue, stealth-stealth] (\xend -0.4,3*\yend/4) -- (\xend + 0.4,3*\yend/4);
\draw[thick, color = black] (0,\yend) -- (\xend,\yend) -- (\xend + \xshift, \yend + \yshift) -- (\xend,\yend) -- (\xend,0);

\draw[thick, color= gray, fill = gray] (\xend/2 - \radd*\cirxshift, \yend/2 + \radd*\ciryshift) -- (\xend/2 + \radd*\cirxshift, \yend/2 - \radd*\ciryshift)  -- (\xend/2 + \radd*\cirxshift + \xshift, \yend/2 - \radd*\ciryshift + \yshift) -- (\xend/2 - \radd*\cirxshift + \xshift, \yend/2 + \radd*\ciryshift + \yshift) -- cycle;
\draw[thick, color = gray, fill = gray] (\xend/2 + \xshift, \yend/2 + \yshift) circle (\radd);
\draw[thick, color = black, fill = black] (\xend/2, \yend/2) circle (\radd);

\draw[color = gray] (0,0) -- (-0.5,\yend/2)node[above, rotate = 90, color = black] {width ($\ell$)} -- (0,\yend);
\draw[color = gray] (0,0) -- (\xend/2,-0.5)node[below,color = black] {length ($L$)} -- (\xend,0);
\draw[color = gray] (\xend,0) -- (\xend + 0.5,-0.5)node[below,color = black, rotate = 0] {height ($h$)} -- (\xend+\xshift,0 + \yshift);
\draw[color = gray,<-] (\xend/2 + \radd*\cirxshift + 0.02, \yend/2 - \radd*\ciryshift - 0.02) -- (\xend/2 + 0.4,\yend/2 - 0.3)node[right, color = black] {radius ($a$)};

\draw[thick, -stealth] (\xend + 1, 0 + \yend/2) -- (\xend + 1, 0.6 + \yend/2)node[right] {$y$};
\draw[thick, -stealth] (\xend + 1, 0 + \yend/2) -- (\xend + 1 + 0.6 ,0 + \yend/2)node[right] {$x$};
\draw[thick, -stealth] (\xend + 1, 0 + \yend/2) -- (\xend  +1 - \xshift*2.5 ,- \yshift*2.5 + \yend/2)node[right] {$z$} ;
\end{tikzpicture}
\caption{Schematic of the narrow channel with cylinder fixed at the center and fluid being forced to oscillate at the channel ends.}
\label{fig:SS_schematic}
\end{figure}

In this paper, we focus on steady streaming in a thin gap between two plates, where the cylinder's axis is orthogonal to the plates, see \cref{fig:SS_schematic} and the experimental setup of \cite{vishwanathan2019steadyViscometry,vishwanathan2019steady}. Here, the walls in the direction of the axis of the cylinder are sufficiently far away, but the plates that are orthogonal to the axis of the cylinder are only two radii apart. For this reason, we do not consider two-dimensional flow, but instead consider planar flow in the direction of the plates while retaining $z$-dependence. With this model, we can capture the effects of the driving flow in a thin gap and produce a steady streaming model that can be efficiently evaluated anywhere within the channel. 


\subsection{Main Results}

In this paper, we develop a three-dimensional planar flow model for steady streaming. The flow is induced by forcing the fluid to oscillate at the ends of the channel. These boundary conditions are implemented through the known series solution for the problem of oscillating flow in an unobstructed rectangular channel. The model includes Fourier series in $t$ and $z$ to reduce the PDEs to two dimensions and an asymptotic expansion to linearize the equations for the stream functions. In our two-dimensional rectangular channel, we solve the resulting equations using the finite element method. Due to the fact that we are solving biharmonic equations, we also use penalty methods to achieve the desired regularity in our solutions. Our model allows us to easily evaluate the steady streaming flow for varying domains, frequencies, and level sets in $z$ throughout the height of the channel. We show that the outer walls do not impact steady streaming near the cylinder as long as these outer walls are sufficiently far away from the cylinder. It is shown that if the outer walls are near to the cylinder then there are qualitative changes to the shapes of the vortices as well as the location of the center of the vortices for different oscillating frequencies. Higher-order solutions are solved for and presented along with the corrected steady streaming.


\subsection{Outline} 
In \cref{sec:Derivation}, we define all the necessary boundary conditions and develop our three-dimensional planar flow model for the steady streaming velocity, including a Fourier series in $z$ to include the confinement effects of the narrow channel. A method to post-process the results is constructed in order to satisfy the no-slip boundary condition on the walls in $z$-direction, which are orthogonal to the axis of the cylinder. In \cref{sec:NumericalApproximation}, we describe and set up the finite element method we use to solve the problem in the narrow channel. We present empirical convergence results for the error due to the discretization in space and the error due to the truncation of the Fourier series. In \cref{sec:results}, we discuss a few different ways our method can be utilized to analyze the steady streaming flow. We evaluate how the steady streaming flow depends on the domain, the frequency, and the location in the height of the channel. Further, we consider the correction term to the steady streaming flow, i.e., the next order solution from the asymptotic expansion, and discuss those results as well. In \cref{sec:discussion}, we conclude with a discussion of the model and the results.


\section{Model Derivation} \label{sec:Derivation}

We begin with the incompressible Navier-Stokes equations in dimensionless form, 
\begin{align}
    \partial_t\u+\ep \u\cdot\nabla\u&=-\nabla p+\frac{1}{\Wo^2}\Delta\u \label{eqn:navstok_nond},\\
    \nabla\cdot\u &= 0. \label{eqn:incom}
\end{align}
Here, $\ep=\frac{s}{a}$ compares the amplitude of the oscillations $s$ to the radius of the cylinder $a$, and $\Wo^2=\frac{a^2 \omega}{\nu}$ is the square Womersley number with $\nu$ being the kinematic viscosity of the fluid and $\omega$ the frequency of the driving oscillations. $\Wo$ compares oscillatory inertial forces to viscous forces. We consider small-amplitude oscillations as compared to the radius of the cylinder such that  $\ep \ll 1$. Finally, all channel dimensions are as in \cref{fig:SS_schematic}, i.e., $\ell$ is the width in the $y$-direction, $L$ is the length in the $x$-direction, and $h$ is the height in the $z$-direction. In writing \cref{eqn:navstok_nond}-\cref{eqn:incom}, we chose the characteristic time scale to be $1/\omega$, the characteristic length scale to be $a$, the characteristic velocity to be $s\omega$, and the characteristic pressure to be $\rho a s\omega^2$, where $\rho$ is the fluid density. 

Next, we introduce the vorticity vector, $\vor=\nabla\times\u$, with the goal of transforming \cref{eqn:navstok_nond,eqn:incom} into a stream function formulation. Taking the curl of \cref{eqn:navstok_nond} gives 
\begin{equation}
\partial_t\vor+\varepsilon(\u\cdot\nabla)\vor=\varepsilon(\vor\cdot\nabla)\u+\frac{1}{\Wo^2}\Delta\vor. \label{eqn:vorticity1}
\end{equation}
At this point, we restrict to three-dimensional planar flow throughout the channel, i.e., $\u$ has the form $\u = u \i + v \j + 0 \k$ where $u=u(x,y,z,t)$ $v=v(x,y,z,t)$. This idea is justified by experiments by Lutz et. al. \cite{lutz2005microscopic}, where planar flow was observed even near the cylinder, as long as the flow was not too close to the $z$-walls. Next, we introduce the stream function $\psi(x,y,z,t)$ such that $u = \partial_y \psi$ and $v = -\partial_x \psi$. This definition for $\psi$, along with the fact that $w=0$, implies that $\u$ trivially satisfies \cref{eqn:incom}. At this point, we now consider $\nabla$ and $\Delta$ to represent the two-dimensional gradient and Laplacian with respect to $x,y$ and we introduce the two-dimensional perpendicular gradient $\gp=\partial_y\i-\partial_x\j$. Substituting these equations into \cref{eqn:vorticity1} and separating the $z$ derivatives from the two-dimensional gradients and Laplacians, we arrive at an equation for $\psi$ alone. Finally, we write the resulting $\i, \j$ components as a single two-dimensional vector equation and the $\k$ component as a scalar equation, which are, respectively

\begin{subequations}
\label{eqn:vorticity2}
\begin{align} 
	&\begin{aligned}
		\partial_t\partial_z\nabla\psi+\varepsilon(\nabla^\perp\psi\cdot\nabla)\partial_z\nabla\psi=&\varepsilon(\partial_z\nabla\psi\cdot\nabla-\Delta\psi\partial_z)\nabla^\perp\psi\\
		&+\frac{1}{\Wo^2}\left(\Delta\partial_z\nabla\psi+\partial_{zzz}\nabla\psi\right), \label{eqn:vorticity2a}
	\end{aligned}\\
	&\partial_t\Delta\psi+\varepsilon(\nabla^\perp\psi\cdot\nabla)\Delta\psi=\frac{1}{\Wo^2}\left(\Delta^2\psi+\partial_{zz}\Delta\psi\right). \label{eqn:vorticity2b}
\end{align}
\end{subequations}
\begin{remark}
We note that after taking a divergence of \cref{eqn:vorticity2a} the nonlinear terms can be factored as a single $z$ derivative such that the entire equation can be integrated with respect to $z$. After setting the constant of integration to 0 one recovers \cref{eqn:vorticity2b}. Thus, keeping both equations in \cref{eqn:vorticity2} is redundant, and we will only consider \cref{eqn:vorticity2b}. 
\end{remark}

\subsection{Boundary conditions}

\begin{figure}[tbhp]
\centering
\begin{tikzpicture}[scale = 0.97]
    \draw[ultra thick, \myred] (0,2.5) node[left]{$y = \frac{\ell}{2a}$} -- (10,2.5);
    \draw[ultra thick, \myblue] (10,2.5) -- (10,0)node[below]{$x = \frac{L}{2a}$};
    \draw[ultra thick, \mygreen] (1,0) arc (0:90:1)node[left]{$x^2 + y^2=1$};
    
    \draw[\myblue] (3.75,1.75) node[above right]{(D) $u = \psi_y = f(y,z)e^{it} + \overline{f(y,z)}e^{-it}$};
    \draw[\myblue] (3.75,1.75) node[below right]{(N) $v = -\psi_x = 0$};

    \draw[\myred] (3.56,3.1) node[above right]{(N) $u= \psi_y = 0$};
    \draw[\myred] (3.56,3.1) node[below right]{(D) $v = -\psi_x = 0$};

    \draw[\mygreen] (1,0.5) node[above right] {(N) $\u \times \unitvect{\nu} = 0$, $\nabla \psi \cdot \unitvect{n} = 0$};
    \draw[\mygreen] (1,0.5) node[below right] {(D)  $\u \cdot \unitvect{\nu} = 0$, $\nabla^{\perp} \psi \cdot \unitvect{n} = 0$};
\end{tikzpicture}
\label{fig:BCschematic}
\caption{Summary of boundary conditions in the $(x,y)-$plane on a quarter of the domain. We have indicated which conditions ultimately are enforced on $\psi$ as Dirichlet (D) and Neumann (N) boundary conditions. The outer rectangular boundary and the cylinder are not to scale.}
\end{figure}
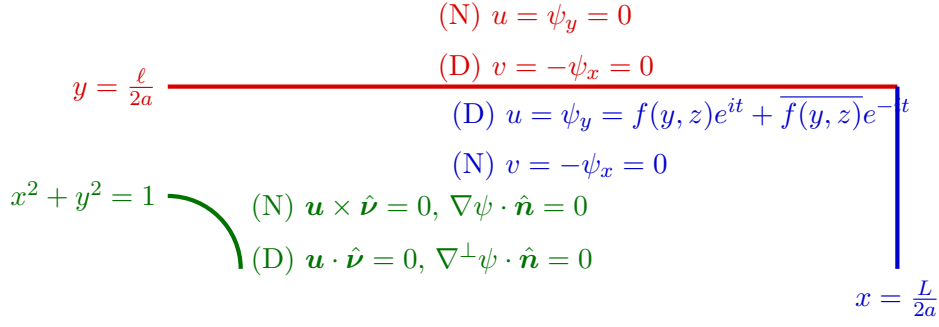

We now carefully construct the boundary conditions which are no-slip and no-penetration on all walls except at the ends of the channel where we impose a velocity flow profile derived from an oscillating pressure in a rectangular channel. We have summarized all the boundary conditions in the $(x,y)$-plane in \cref{fig:BCschematic}. Exploiting the symmetry of the domain we have only illustrated this on a quarter of the boundary. In what follows, we set up up the boundary conditions for the full three-dimensional domain, but our final model is two-dimensional, in $(x,y)$, therefore we only illustrate those boundary conditions in \cref{fig:BCschematic}.

On the stationary walls, we have no-slip, which can be summarized as $\u \times \unitvect{\nu} = - \partial_x \psi \nu_3 \i - \partial_y \psi \nu_3 \j + (\partial_x \psi \nu_1 + \partial_y \psi \nu_2) \k= 0$ , where $\unitvect{\nu}$ is the three-dimensional unit normal to the surface. All surfaces in our domain, see \cref{fig:SS_schematic}, are either parallel to $\k$ or orthogonal to $\k$. As a result, $\nu_3=0$ on $x=\pm\frac{L}{2a}$, $y=\pm\frac{\ell}{2a}$, and on the cylinder $x^2+y^2=1$ and we therefore have a zero Neumann boundary condition of the form $\nabla\psi \cdot \unitvect{n}=0$, where $\unitvect{n}$ is the two-dimensional unit vector at a fixed $z$-level. This is indicated by the (N) in \cref{fig:BCschematic}. On the $z=\pm\frac{h}{2a}$ walls, where $\nu_1 = 0 =\nu_2$, the no-slip condition yields $\partial_x \psi = 0 = \partial_y \psi$ which is equivalent to requiring $\psi$ to be constant on these boundaries. 

Next, we note that the no-penetration condition is trivially satisfied on the $z = \pm \frac{h}{2a}$ walls, since $w=0$. The no-penetration conditions at the walls $y=\pm\frac{\ell}{2a}$ and the cylinder are $\u \cdot \unitvect{\nu} = 0$ and in terms of the stream function this is $\gp \psi \cdot \unitvect{n} = 0$. We can instead write this as $\nabla \psi \cdot \unitvect{n}^{\perp} = 0$, where $\unitvect{n}^{\perp}$ is trivially defined since $\unitvect{n}$ is two-dimensional. Therefore, the directional derivative of $\psi$ in the tangential direction of these boundaries is 0, i.e., $\psi$ is constant along level sets in $z$ on the $y = \pm \frac{\ell}{2a}$ walls and the cylinder. This is indicated by the (D) on the top walls and on the cylinder in \cref{fig:BCschematic}.

\begin{figure}[tb]
\centering
	\subfloat[Real part of $f(y,z)$.]{\label{fig:BoundaryFunctionReal}
	\includegraphics[width = 0.99\textwidth,trim = {1.2cm 0.6cm 8cm 0.6cm},clip=true]{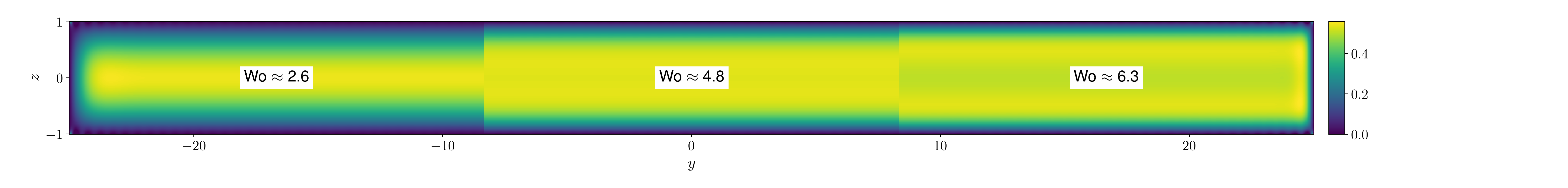}}
 
 	\subfloat[Imaginary part of $f(y,z)$.]{\label{fig:BoundaryFunctionImag}
	\includegraphics[width = 0.99\textwidth,trim = {1.2cm 0.6cm 8cm 0.6cm},clip=true]{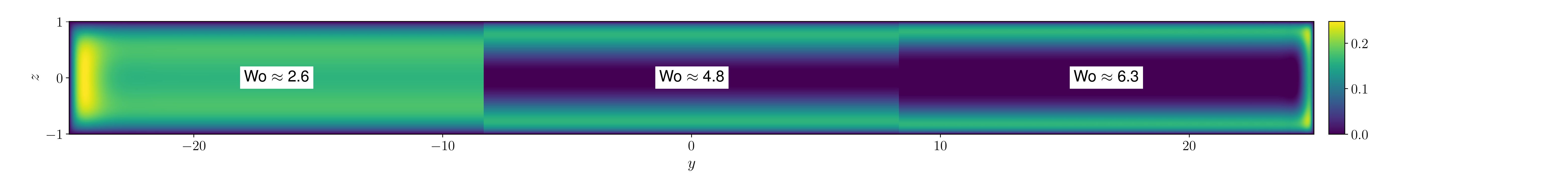}}
\caption{Real and imaginary parts of the series solution $f(y,z)$, truncated at $M=100$. We take advantage of the symmetry across both the $y$ and $z$ axes and the nearly constant nature for the interior $y$ values and plot $f(y,z)$ for $\Wo \approx 2.6$ on the left third of the domain, $\Wo \approx 4.8$ on the middle third of the domain, and $\Wo \approx 6.3$ on the last third of the domain.}
\label{fig:BoundaryFunction}
\end{figure}

To determine the oscillatory flow boundary conditions at the channel ends, $x = \pm \frac{L}{2a}$, we temporarily neglect the cylinder. We then solve the auxiliary problem of unidirectional oscillatory flow in a rectangular channel driven by a dimensionless pressure gradient that oscillates in time, of the form $ \sin(t)$, see \cite{moore2015theory,o1975pulsatile} for the detailed calculation. The resulting velocity profile is given by $f(y,z) e^{it} + \overline{f(y,z)} e^{-it}$, where
\begin{equation*}
    \begin{split}
        f(y,z)=\frac{1}{2}-\frac{2}{\pi}\sum_{m\geq 0}\frac{(-1)^m}{2m+1}
        &\left[\cos\left(\frac{(2m+1)\pi az}{h}\right)\frac{\cosh\left(\sqrt{\lambda_2}y\right)}{\cosh\left(\frac{\sqrt{\lambda_2} \ell}{2a}\right)}\right.\\
        &\left.+\cos\left(\frac{(2m+1)\pi ay}{\ell}\right)\frac{\cosh\left(\sqrt{\lambda_1}z\right)}{\cosh\left(\frac{\sqrt{\lambda_1}h}{2a}\right)}\right],
    \end{split}
\end{equation*}
 with $\lambda_1=i\Wo^2+\left(\frac{(2m+1)\pi a}{\ell}\right)^2$ and $\lambda_2=i\Wo^2+\left(\frac{(2m+1)\pi a}{h}\right)^2$. The real and imaginary parts of $f(y,z)$ are shown in \cref{fig:BoundaryFunction}.
Then we set the oscillatory boundary condition as  $u =\partial_y\psi= f(y,z) e^{it} + \overline{f(y,z)} e^{-it}$ at $x=\pm\frac{L}{2a}$. Since this derivative on $\psi$ is tangential to the boundary, we will integrate this condition and enforce it as a Dirichlet condition on $\psi$, as indicated by the (D) in \cref{fig:BCschematic}. Therefore, we define $F(y,z)=\int f(y,z)~dy$ to be the anti-derivative of $f(y,z)$ with respect to $y$, such that
 \begin{equation}
     \begin{split}
        F(y,z)=\frac{1}{2}y-\frac{2}{\pi}\sum_{m\geq 0}&\frac{(-1)^m}{2m+1}\left[\cos\left(\frac{(2m+1)\pi az}{h}\right)\frac{\sinh\left(\sqrt{\lambda_2}y\right)}{\sqrt{\lambda_2}\cosh\left(\frac{\sqrt{\lambda_2} \ell}{2a}\right)}\right.\\
&\left.+\sin\left(\frac{(2m+1)\pi ay}{\ell}\right)\frac{\ell}{(2m+1)\pi a}\frac{\cosh\left(\sqrt{\lambda_1}z\right)}{\cosh\left(\frac{\sqrt{\lambda_1}h}{2a}\right)}\right].
     \end{split}
     \label{eqn:F_z}
 \end{equation} 
 We have omitted the constant of integration for simplicity, but this will be chosen to be zero and is discussed in detail next. 
 
 Each of the Dirichlet conditions has an arbitrary constant (or arbitrary function of $z$). Since we only need to determine $\psi$ up to an arbitrary constant we let $\psi=0$ on the cylinder and trace out the constants for the boundary conditions on all other surfaces. Requiring that $\psi$ is continuous we determine that all the constants must be zero. The value of $\psi$ at the intersection between the $x$ and $y$ walls automatically gives the constant value of $\psi$ along the $y$-walls for each fixed $z$. 
 
 In summary, the boundary conditions are
\begin{align}
\psi &= \left(F(y,z) e^{it} + \overline{F(y,z)} e^{-it}\right) , \quad  \text{ on } x = \pm \frac{L}{2a}, \text{ and  on }  y= \pm \frac{\ell}{2a}, \label{eqn:BCendsNDStream} \\
\psi &= 0, \quad \text{ on }x^2+y^2=1,\text{ and on } z = \pm \frac{h}{2a}, \label{eqn:BCzeroNDStream} \\
\partial_{\unitvect{n}} \psi &= 0, \quad \text{ on } x=\pm\frac{L}{2a}, \text{ on } y=\pm\frac{\ell}{2a}, \text{ and on }x^2+y^2=1,  \label{eqn:BCNeu}
\end{align}
where $\partial_{\unitvect{n}}$ denotes the normal derivative along the boundary for a fixed level in $z$.

\subsection{Fourier and asymptotic expansions}
To separate the steady and unsteady parts as well as the primary and secondary flow, we perform Fourier series and an asymptotic expansion in $\psi$. As there will be three separate expansions,  we will use subscripts to represent Fourier indices and superscripts to represent orders in the asymptotic expansion. 

Beginning with a Fourier series in time, we consider  $\psi=\displaystyle\sum_{n=-\infty}^\infty\psi_n(x,y,z) e^{int}$, and we require that $\psi_{-n} = \overline{\psi}_n$ since the stream function is real. Plugging in \cref{eqn:vorticity2b}, using the Cauchy product for the nonlinear terms, and then collecting terms in $e^{int}$, we have 
\begin{equation}
	in\Delta\psi_n+\varepsilon\sum_{m=-\infty}^\infty\nabla^\perp\psi_m\cdot\nabla\Delta\psi_{n-m}=\frac{1}{\Wo^2}\left(\Delta^2\psi_n+\partial_{zz}\Delta\psi_n\right). \label{eqn:AfterTime}
\end{equation}
From \cref{eqn:BCendsNDStream,eqn:BCzeroNDStream,eqn:BCNeu}, it is obvious that the only nonzero boundary condition is
\begin{equation}
    \psi_{1} =  F(y,z), \quad \psi_{-1} =  \overline{F(y,z)} , \quad \text{ on } x = \pm \frac{L}{2a} \text{ and on } y= \pm \frac{\ell}{2a}. \label{eqn:BCendsFourTimen1}
\end{equation}

Next, we proceed in a similar fashion with a Fourier series in $z$, and consider the expansion $\psi_n=\displaystyle\sum_{k=-\infty}^\infty\psi_{n,k}(x,y)e^{i\frac{2\pi a}{h}kz}$. Substituting into \cref{eqn:AfterTime}, using the Cauchy product on the nonlinear terms, and collecting $e^{i\frac{2\pi a}{h}kz}$ terms, we find
\begin{equation}
		i n\Delta\psi_{n,k}+\varepsilon\sum_{m=-\infty}^\infty\sum_{j=-\infty}^\infty\nabla^\perp\psi_{m,j}\cdot\nabla\Delta\psi_{n-m,k-j}
		=\frac{1}{\Wo^2}\left(\Delta^2\psi_{n,k}-\alpha_k \Delta\psi_{n,k}\right) \label{eqn:AfterZ},
\end{equation}
where $\alpha_k=\left(\frac{2\pi a k}{h}\right)^2$. The zero Dirichlet and zero Neumann boundary conditions, \cref{eqn:BCzeroNDStream,eqn:BCNeu}, easily translate to the $\psi_{n,k}$ coefficients as 
\begin{align}
\psi_{n,k} &= 0, \quad \text{ on } x^2+y^2=1, \text{ for all } n,k \label{eqn:BCDirFourZ} \\
\partial_{\unitvect{n}} \psi_{n,k} &= 0, \quad \text{ on all } (x,y) \text{ boundaries}, \text{ for all } n, k.\label{eqn:BCNeuFourZ}
\end{align}

\begin{remark}
To satisfy  no-slip \cref{eqn:BCzeroNDStream} at $z=\pm\frac{h}{2a}$, we must have $\psi_n=0$ at $z=\pm\frac{h}{2a}$ for all $n$ and thus $\sum_{k=-\infty}^\infty(-1)^k\psi_{n,k}=0$ for all $n$. In practice, we truncate the infinite series and will not enforce this boundary condition. In  \cref{sec:slip}, we discuss a procedure to correct for the non-zero slip at the $z$ walls.
\end{remark}

For $n\neq\pm1$, \cref{eqn:BCendsFourTimen1} straightforwardly gives 
\begin{equation}
    \psi_{n,k} = 0, \quad \text{ on } x = \pm \frac{L}{2a},\text{ and on } y= \pm \frac{\ell}{2a}, \text{ for }n \neq \pm 1. \label{eqn:BCendsFourZ0} 
\end{equation}
To derive the boundary conditions at the channel ends when $n=\pm 1$, we consider the Fourier expansion $\ds F(y,z) = \sum_{k=-\infty}^{\infty} \widehat{F}_k(y) e^{i\frac{2\pi a}{h}kz}$, where $F(y,z)$ is defined in \cref{eqn:F_z} and 
$$\ds \widehat{F}_k(y) = \frac{a}{h} \int_{-\frac{h}{2a}}^{\frac{h}{2a}} F(y,z) e^{-i\frac{2\pi a}{h}kz} ~ dz.$$
These $\widehat{F}_k(y)$ coefficients are solved with Mathematica \cite{Mathematica}. It follows from \cref{eqn:BCendsFourTimen1} that the boundary condition on $x = \pm \frac{L}{2a}$ and $ y= \pm \frac{\ell}{2a}$ is
\begin{equation}
    \psi_{\pm 1,k}= \widehat{F}_k(y) = \begin{dcases} 
    \frac{8}{\pi} (-1)^k \sum_{m \geq 0} a_k^m(y) & \text{if }k \neq 0, \\
    \frac{1}{2}y + \frac{4}{\pi} (-1)^k \sum_{m \geq 0} a_k^m(y) & \text{if }k = 0,
    \end{dcases}
    \label{eqn:BCendsFourZ}
\end{equation}
where, 
\begin{align*}
    a_k^m(y) &= \frac{1}{\pi}\frac{1}{4k^2-(2m+1)^2}\frac{\sinh(\sqrt{\lambda_2}y)}{\sqrt{\lambda_2}\cosh\left(\sqrt{\lambda_2}\frac{\ell}{2a}\right)} \\
    & \quad \quad - \frac{ah(-1)^m\sqrt{\lambda_1}}{4a^2\pi^2k^2+h^2\lambda_1}\frac{\ell}{(2m+1)^2\pi a} \tanh\left(\sqrt{\lambda_1}\frac{h}{2a}\right)\sin\left(\frac{(2m+1)\pi ay}{\ell}\right).
\end{align*}
In practice, we chose to truncate the $\widehat{F}_k(y)$ coefficients at $M=100$. We tested multiple values of $M$ and in particular tested $M=50$ and $M=200$. The boundary function $f(y,z)$ for the prescribed oscillating flow appeared to be well resolved at each of these values and the solutions near the cylinder had no significant change between these values of $M$. All results presented in \cref{sec:results} were computed with $M=100$.

Finally, we asymptotically expand the Fourier coefficients in $\ep$ as $\psi_{n,k}=\psi_{n,k}^{(0)}+\varepsilon\psi_{n,k}^{(1)} + \cdots$ and begin with the finite approximation, $\psi_{n,k}\approx \psi_{n,k}^{(0)}+\varepsilon\psi_{n,k}^{(1)}$. Substituting this approximation into \cref{eqn:AfterZ} we collect the $O(\ep^0)$ and $O(\ep^1)$ terms, which are, respectively,
\begin{align}
	\Delta^2\psi_{n,k}^{(0)}-\left(in\Wo^2 + \alpha_k \right)\Delta\psi_{n,k}^{(0)}&=0\label{eqn:zero_order},\\
	\Delta^2\psi_{n,k}^{(1)}-\left(in\Wo^2+\alpha_k\right)\Delta\psi_{n,k}^{(1)}&=\Wo^2\sum_{m=-\infty}^\infty\sum_{j=-\infty}^\infty\nabla^\perp\psi_{m,j}^{(0)}\cdot\nabla\Delta\psi_{n-m,k-j}^{(0)}.\label{eqn:first_order}
\end{align}
 We point out that evaluating \cref{eqn:zero_order,eqn:first_order,eqn:BCendsFourZ0,eqn:BCendsFourZ,eqn:BCDirFourZ,eqn:BCNeuFourZ} at $-n$  results in the conjugate equation for each and it follows that $\psi_{-n} = \overline{\psi_{n}}$, which is necessary for $\psi$ to be real, as previously discussed. Furthermore, since \cref{eqn:zero_order} is homogeneous and \cref{eqn:BCendsFourZ0,eqn:BCDirFourZ,eqn:BCNeuFourZ} are homogeneous for $n\neq \pm 1$ we have that $\psi_n = 0$ for $n\neq \pm1$. Therefore, we only need to solve \cref{eqn:zero_order} for $n=1$. Recalling that the steady streaming solution is the time-independent component of the secondary flow we now formalize this derivation with the following definition.

\begin{definition} \label{def:SS}
The steady streaming solution is $\psi_0^{(1)} = \displaystyle\sum_{k=-\infty}^\infty\psi_{0,k}^{(1)}(x,y)e^{i\frac{2\pi a}{h}kz},$
where for each $k$, $\psi_{0,k}^{(1)}$ satisfies zero Dirichlet and zero Neumann boundary conditions on all boundaries and is a solution to 
\begin{equation}	
\Delta^2\psi_{0,k}^{(1)}-\alpha_k\Delta\psi_{0,k}^{(1)}=\Wo^2 \sum_{j=-\infty}^\infty \nabla^\perp\psi_{1,j}^{(0)}\cdot\nabla\Delta\overline{\psi_{1,k-j}^{(0)}} + \nabla^\perp\overline{\psi_{1,j}^{(0)}}\cdot\nabla\Delta\psi_{1,k-j}^{(0)}. \label{eqn:NewSS}
\end{equation}
It is assumed that the lower order solution $\psi_{1,k}^{(0)}$ is known and satisfies \cref{eqn:zero_order,eqn:BCendsFourZ,eqn:BCNeuFourZ,eqn:BCDirFourZ,eqn:BCendsFourZ0}  with $n=1$.
\end{definition}
\begin{remark}
Since $\alpha_k = \alpha_{-k}$ it is obvious from \cref{eqn:zero_order,eqn:BCendsFourZ,eqn:BCendsFourZ0,eqn:BCDirFourZ,eqn:BCNeuFourZ} that $\psi_{1,k}^{(0)} = \psi_{1,-k}^{(0)}$. We now evaluate \cref{eqn:NewSS} for $k \to -k$ and show that the right hand side is the same as when considering $k$. We only need to consider $\nabla^\perp\psi_{1,j}^{(0)}\cdot\nabla\Delta\overline{\psi_{1,-k-j}^{(0)}}$ and the other term follows by the same logic. With a change of variables $j \to -j$ we get $\nabla^\perp\psi_{1,-j}^{(0)}\cdot\nabla\Delta\overline{\psi_{1,-k+j}^{(0)}}$ and the result follows from the symmetry of the lower order solutions. Therefore, we have that $\psi_{0,-k}^{(1)} = \psi_{0,k}^{(1)}$.
\end{remark}


\subsection{Post-processing slip at the vertical walls} \label{sec:slip}

In order to satisfy the no-slip boundary conditions on the walls in $z$ direction we post-process all solutions and construct this procedure here. Assuming all the $\psi_{n,k}^{(i)}$ are known and satisfy \cref{eqn:zero_order,eqn:NewSS}, we start by defining $\ds \widetilde{\psi}_{n,k}^{(i)} = \psi_{n,k}^{(i)} - (-1)^k\beta_k \sum_j (-1)^j\psi_{n,j}^{(i)}$, where $\beta_k \geq 0$ for all $k$ and $\ds \sum_k \beta_k = 1$. We then define $\ds \widetilde{\psi}_n^{(i)}(x,y,z) =\sum_{k} \widetilde{\psi}_{n,k}^{(i)}(x,y) e^{i\frac{2\pi a}{h}kz}$. Evaluating at $z=\pm\frac{h}{2a}$ it follows by direct computation and the fact that the $\beta_k$ sum to one that $\widetilde{\psi}_n^{(i)}=0$, i.e., satisfies the no-slip boundary condition, on the $z$-walls.

\begin{figure}[tbhp]
\centering
	\subfloat[\\ $K=9$]{\label{fig:slipz_k9}
	\includegraphics[width = 0.24\textwidth,trim = {1cm 0.9cm 1.75cm 1.5cm},clip=true]{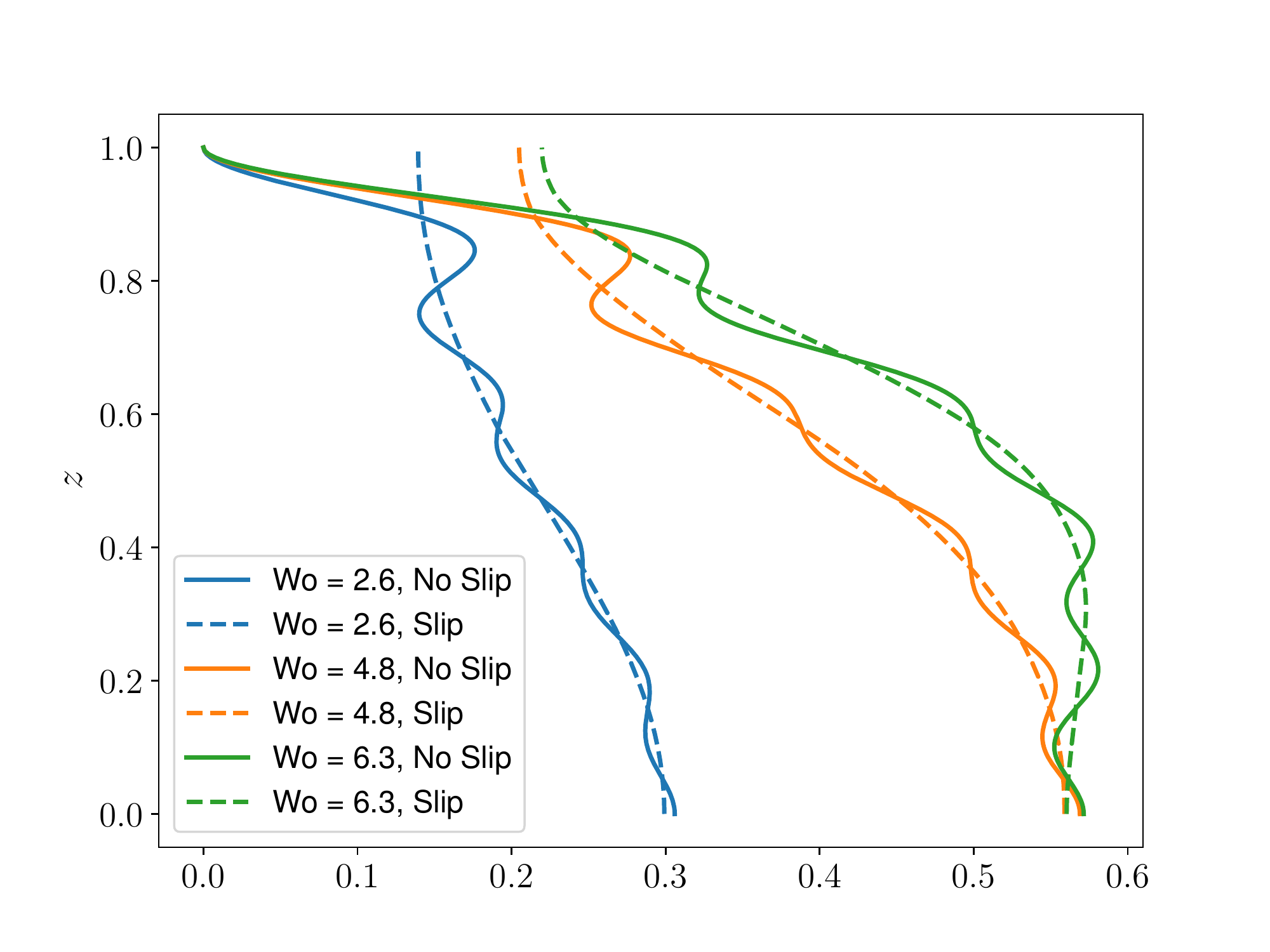}}
 	\subfloat[\\ ${K=10}$]{\label{fig:slipz_k10}
	\includegraphics[width = 0.24\textwidth,trim = {1cm 0.9cm 1.75cm 1.5cm},clip=true]{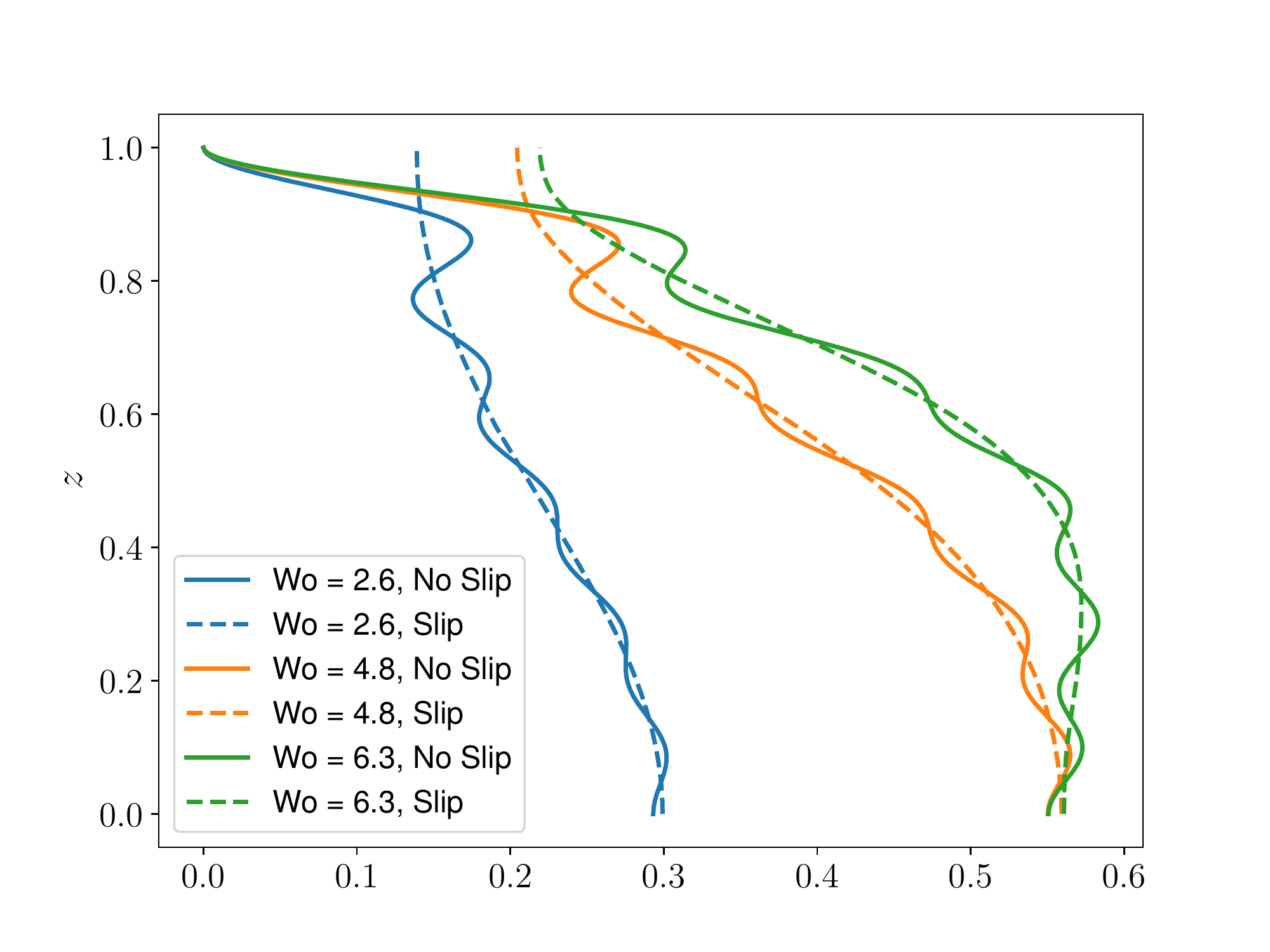}}
 	\subfloat[\\ $K=11$]{\label{fig:slipz_k11}
	\includegraphics[width = 0.24\textwidth,trim = {1cm 0.9cm 1.75cm 1.5cm},clip=true]{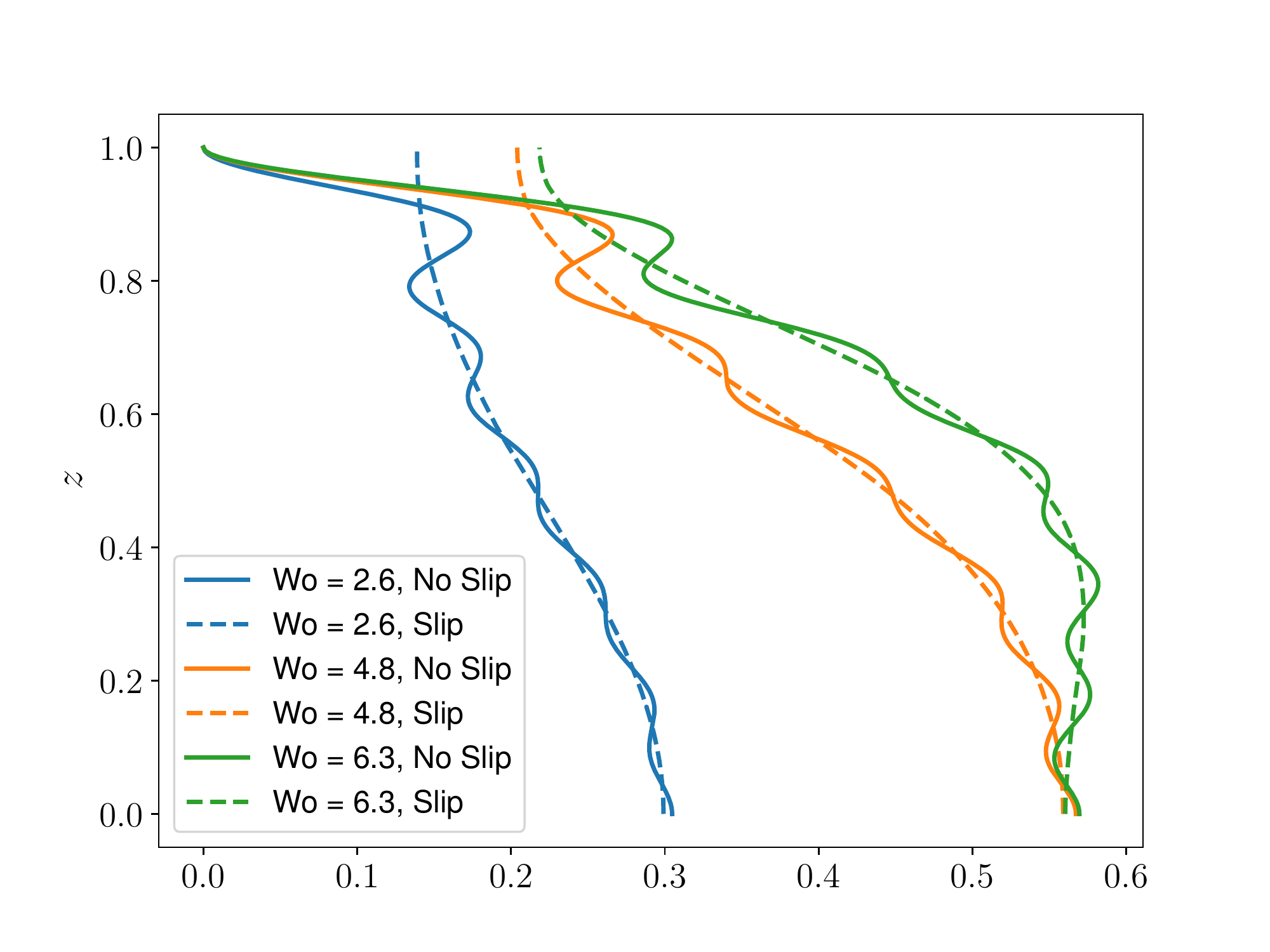}}
 	\subfloat[\\ $K=12$]{\label{fig:slipz_n3_k12}
	\includegraphics[width = 0.24\textwidth,trim = {1cm 0.9cm 1.75cm 1.5cm},clip=true]{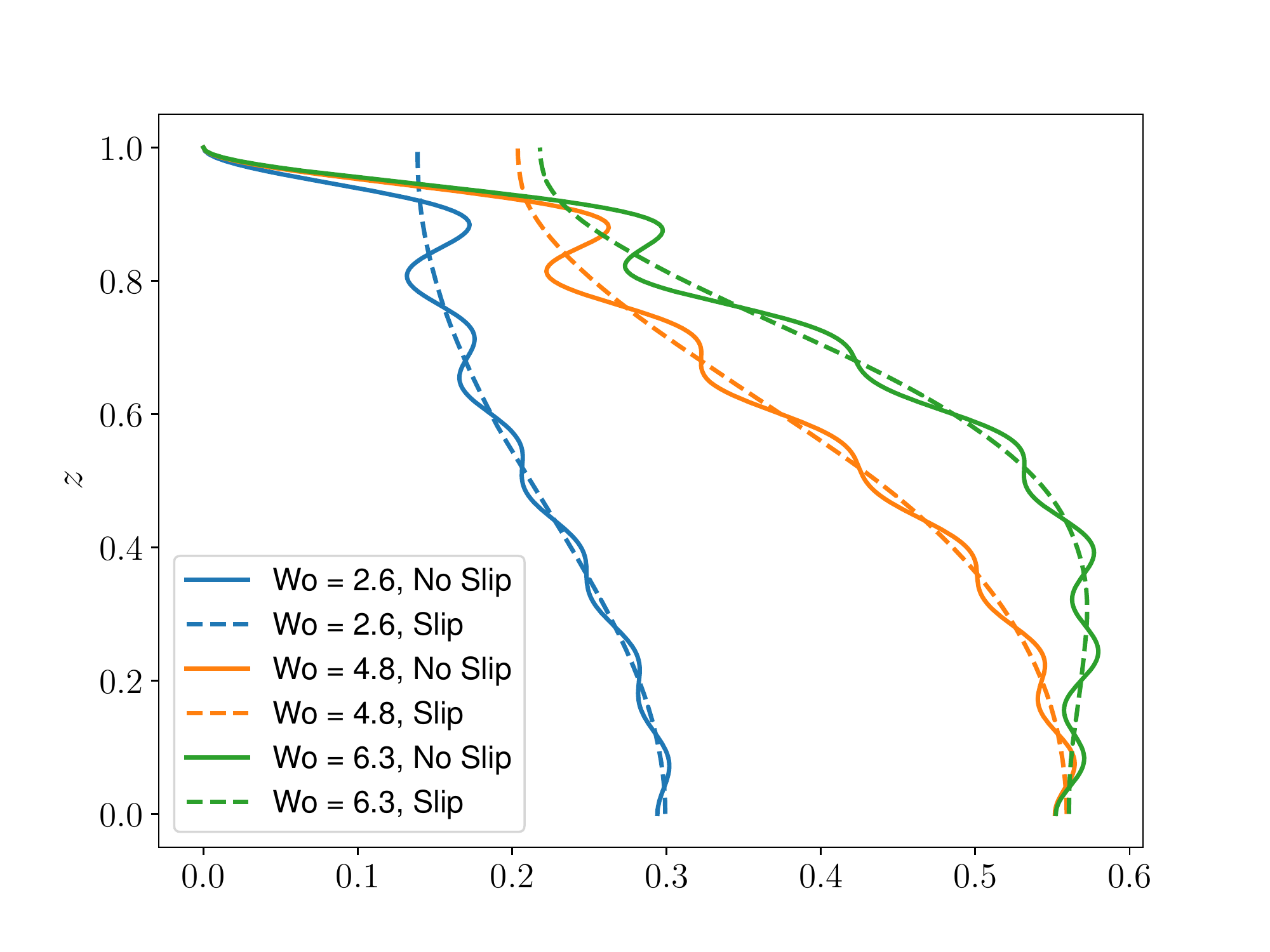}}
\caption{Plotting the maximum steady streaming velocity, for level sets in $z$, to visualize how well the $\widetilde{\psi}_0^{(1)}$ (solid lines) approximates $\psi_0^{(1)}$ (dashed lines). Solutions are plotted for $\Wo \approx 2.6, 4.8, 6.3$. Plots (a)-(d) are for $K=9,\ldots,12$, respectively, where $K$ is the truncation for the $z$-Fourier series. As the solutions are symmetric in $z$ we have only plotted the solutions in the upper half of the domain, i.e., $z \in [0,\frac{h}{2a}]$.}
\label{fig:slipz_vsn2}
\end{figure}

In \cref{fig:slipz_vsn2}, we have plotted the maximum velocity, for level sets in $z$, of the no-slip steady streaming solution and the slip steady streaming solution, for $\Wo\approx 2.6, 4.8, 6.3$ and $K=9,10,11,12$, where $K$ is the maximal index corresponding to the truncation of the $z$-Fourier series, i.e. $k\in[-K,K]$. For all the results presented in \cref{sec:results} $K=10$. In \cref{fig:slipz_vsn2}, we chose these values of $K$ to illustrate the fact that the no-slip curve alternates across the slip curve for even versus odd values of $K$ at the midheight of the channel, i.e., $z=0$. This observation is important when considering the convergence in the $z$-Fourier modes, which will be discussed in \cref{sec:NumericalConvergence}. The $\Wo$ has little effect on how well $\widetilde{\psi}_0^{(1)}$ approximates $\psi_0^{(1)}$. We see that the solution where no-slip is enforced well approximates the slipping solution throughout $z\in (-\frac{h}{2a},\frac{h}{2a})$, although oscillations have been introduced. It is obvious from \cref{fig:slipz_vsn2} that the price to be paid for enforcing no-slip in this way is the introduction of spurious oscillations in our solutions. As expected, as we increase $K$ we see that $\widetilde{\psi}_0^{(1)}$ better approximates $\psi_0^{(1)}$ for $z\in (-\frac{h}{2a},\frac{h}{2a})$. In particular, we compare the flow profiles in \cref{fig:slipz_vsn2} to the experiments in \cite{lutz2005microscopic} and note the flattening of the profile in the middle of the channel for increasing $\Wo$ which is consistent with those experiments. Finally, we observe a clear relation with respect to the $z$ dependence between the slip and no-slip solutions in \cref{fig:slipz_vsn2} with the driving profile $f(y,z)$, shown in \cref{fig:BoundaryFunction}. 

For the remainder of this paper, the $\ \widetilde{}\ $ will be dropped, but all solutions will be post-processed to be of the form $\widetilde{\psi}_n^{(i)}$. That is, all solutions, for each order of the asymptotic expansion, will satisfy no-slip on the $z = \pm\frac{h}{2a}$.


\subsection{Higher Order Solution}
We first consider which Fourier coefficients are non-zero as we consider higher order asymptotic solutions, which we have summarized in \cref{fig:time_schematic}, where we have indicated the steady streaming components with red text. Therefore, the next order solution for the steady streaming equation is actually two orders away and of the form $\psi_{0}^{(1)} + \ep^2 \psi_{0}^{(3)}.$ 

\begin{figure}[tbhp]
\centering
\begin{tikzpicture}[scale = 1,  every node/.style={scale=0.9}];
\draw[ultra thick](1,0)--(1,5);
\draw[thick](2,0)--(2,5);
\draw[thick](3,0)--(3,5);
\draw[thick](4,0)--(4,5);
\draw[thick](5,0)--(5,5);
\draw[thick](6,0)--(6,5);
\draw[thick](7,0)--(7,5);
\draw[thick](8,0)--(8,5);
\draw[thick](9,0)--(9,5);
\draw[thick](10,0)--(10,5);

\draw[ultra thick](0,0)--(10,0);
\draw[thick](0,1)--(10,1);
\draw[thick](0,2)--(10,2);
\draw[thick](0,3)--(10,3);
\draw[ultra thick](0,4)--(10,4);

\draw[] (1.5,4.5)node[rotate = 45] {$n=-4$};
\draw[] (2.5,4.5)node[rotate = 45] {$n=-3$};
\draw[] (3.5,4.5)node[rotate = 45] {$n=-2$};
\draw[] (4.5,4.5)node[rotate = 45] {$n=-1$};
\draw[] (5.5,4.5)node[rotate = 45,color = red] {$n=0$};
\draw[] (6.5,4.5)node[rotate = 45] {$n=1$};
\draw[] (7.5,4.5)node[rotate = 45] {$n=2$};
\draw[] (8.5,4.5)node[rotate = 45] {$n=3$};
\draw[] (9.5,4.5)node[rotate = 45] {$n=4$};

\draw[] (0.5,3.5)node {$O\left(\epsilon^0\right)$};
\draw[] (0.5,2.5)node {$O\left(\epsilon^1\right)$};
\draw[] (0.5,1.5)node {$O\left(\epsilon^2\right)$};
\draw[] (0.5,0.5)node {$O\left(\epsilon^3\right)$};

\draw[] (4.5,3.5)node {$\psi_{-1}^{(0)}$};
\draw[] (6.5,3.5)node {$\psi_{1}^{(0)}$};

\draw[] (3.5,2.5)node {$\psi_{-2}^{(1)}$};
\draw[] (5.5,2.5)node[color =red] {$\psi_{0}^{(1)}$};
\draw[] (7.5,2.5)node {$\psi_{2}^{(1)}$};

\draw[] (2.5,1.5)node {$\psi_{-3}^{(2)}$};
\draw[] (4.5,1.5)node {$\psi_{-1}^{(2)}$};
\draw[] (6.5,1.5)node {$\psi_{1}^{(2)}$};
\draw[] (8.5,1.5)node {$\psi_{3}^{(2)}$};

\draw[] (1.5,0.5)node {$\psi_{-4}^{(3)}$};
\draw[] (3.5,0.5)node {$\psi_{-2}^{(3)}$};
\draw[] (5.5,0.5)node[color = red] {$\psi_{0}^{(3)}$};
\draw[] (7.5,0.5)node {$\psi_{2}^{(3)}$};
\draw[] (9.5,0.5)node {$\psi_{4}^{(3)}$};

\draw[color = blue, ultra thick, dashed] (5.5,0)--(1.75,3.75);
\draw[color = blue, ultra thick, dashed] (5.5,0)--(9.25,3.75);

\draw[color = green!50!black, ultra thick, dashed] (5.5,2)--(3.75,3.75);
\draw[color = green!50!black, ultra thick, dashed] (5.5,2)--(7.25,3.75);

\end{tikzpicture}
\caption{Schematic showing which Fourier coefficients in time are non-zero at each asymptotic order and how the higher solutions depend on the lower order solutions. Blank squares indicate the coefficient is zero there. The dashed lines illustrate the ``cone of dependence'', indicating that the function at the point of the cone depends on all the functions above the dashed lines.}
\label{fig:time_schematic}
\end{figure} 

From \cref{fig:time_schematic}, we can determine which lower-order solutions any function depends on. For any function, as one goes up one level in the schematic, they also go out when level. Tracing back to $O(\epsilon^0)$, all the functions needed to compute the original function can be determined. We call this the cone of dependence, and it is illustrated with the green dashed lines for $\psi_0^{(1)}$, and the blue dashed lines for $\psi_0^{(3)}$. Therefore, since the equations for $\psi_1^{(0)}$ and $\psi_0^{(1)}$ have already been determined, we only need to construct the PDEs for $\psi_2^{(1)}$, $\psi_1^{(2)}$, and $\psi_0^{(3)}$. All boundary conditions for all solutions other than $\psi_{1,k}^{(0)}$ are zero Dirichlet and zero Neumann. We omit the details of the similar derivation for these higher-order equations, which are

\begin{align}
    &\begin{aligned}
        \Delta^2\psi_{2,k}^{(1)}-\left(2i\Wo^2 + \alpha_k \right)\Delta\psi_{2,k}^{(1)} = \Wo^2 \sum_{j=\infty}^{\infty} \gp \psi_{1,j}^{(0)} \cdot \nabla \Delta \psi_{1,k-j}^{(0)}
        \label{eqn:psi_12}
    \end{aligned}\\
    &\begin{aligned}
        \Delta^2\psi_{1,k}^{(2)} &- \left(i\Wo^2 + \alpha_k \right)\Delta\psi_{1,k}^{(2)} =  \Wo^2 \sum_{j=\infty}^{\infty}\left( \gp \psi_{2,j}^{(1)} \cdot \nabla \Delta \overline{\psi_{1,k-j}^{(0)}} \right. \\ 
        & \left. + \gp \psi_{0,j}^{(1)} \cdot \nabla \Delta \psi_{1,k-j}^{(0)} + \gp \overline{\psi_{1,j}^{(0)}} \cdot \nabla \Delta \psi_{2,k-j}^{(1)} + \gp \psi_{1,j}^{(0)} \cdot \nabla \Delta \psi_{0,k-j}^{(1)} \right)
        \label{eqn:psi_21}
    \end{aligned}
\end{align} 
Finally, the correction term to the steady steaming solution satisfies 
\begin{equation}
    \begin{aligned} 
        \Delta^2\psi_{0,k}^{(3)} &- \alpha_k\Delta\psi_{0,k}^{(3)} =  \Wo^2 \sum_{j=\infty}^{\infty}\bigg(\gp \overline{\psi_{1,j}^{(2)}} \cdot \nabla \Delta \psi_{1,k-j}^{(0)} + 
        \gp \psi_{1,j}^{(2)} \cdot \nabla \Delta  \overline{\psi_{1,k-j}^{(0)}}\\
        &+
        \gp \psi_{2,j}^{(1)} \cdot \nabla  \Delta \overline{\psi_{2,k-j}^{(1)}} + 
        \gp \psi_{0,j}^{(1)} \cdot \nabla \Delta \psi_{0,k-j}^{(1)} 
         + \gp \overline{ \psi_{2,j}^{(1)}} \cdot \nabla \Delta \psi_{2,k-j}^{(1)} \\
        &+
        \gp \overline{\psi_{1,j}^{(0)}} \cdot \nabla \Delta \psi_{1,k-j}^{(2)} + 
        \gp \psi_{1,j}^{(0)} \cdot \nabla \Delta  \overline{\psi_{1,k-j}^{(2)}}  \bigg) 
    \end{aligned}
    \label{eqn:psi_30}
\end{equation}


\section{Numerical Approximation} \label{sec:NumericalApproximation}

We solve \cref{eqn:NewSS} numerically via the finite element method and specifically use FEniCS \cite{LoggEtal102012} (version 2019.1.0) via Python (version 3.9.1) to do so. We generate our meshes via the \texttt{distmesh} package \cite{persson2004simple} via Matlab. We then transfer the meshes to FEniCS, and all the remaining work is completed in Python. We construct symmetric meshes by first creating a mesh for $D = \left\{(x,y) \in \R^2 :\ x \in [0,\frac{L}{2a}], \ y \in [0,\frac{\ell}{2a}], \text{ and } x^2+y^2 \geq 1 \right\}$ via \texttt{distmesh}. Then we mirror $D$ across the $x$ axis, then the $y$ axis, remove any duplicate points, and create a new Delaunay triangulation for the final mesh for $\Omega = \left\{(x,y) \in \R^2 :\ x \in [-\frac{L}{2a}\right.$, $\frac{L}{2a}],$  $\left.y \in [-\frac{\ell}{2a},\frac{\ell}{2a}], \text{ and } x^2+y^2 \geq 1 \right\}$. 

To implement a PDE with complex coefficients into FEniCS \cite{LoggEtal102012} it is necessary to separate real and imaginary parts and instead implement a system of real-valued PDEs. We note that to solve \cref{eqn:NewSS} it is only necessary to separate the $O(\ep^0)$ coefficents as the steady streaming solution is purely real. Therefore, before considering the weak formulation, we will pre-process \cref{eqn:zero_order,eqn:NewSS} to remove any complex valued constants or functions.

We start by expanding the lower order solution as $\psi_{1,k}^{(0)} = \zeta_{1,k}^{(0)}+i \xi_{1,k}^{(0)}$ where $\zeta_{1,k}^{(0)}$ and $\xi_{1,k}^{(0)}$ are real functions. After substituting into \cref{eqn:zero_order}, with $n=1$, and separating real and imaginary terms, we have
\begin{subequations}
	\begin{align}
		&\Delta^2\zeta_{1,k}^{(0)}-\alpha_k\Delta\zeta_{n,k}^{0}+\Wo^2\Delta\xi_{1,k}^{0}=0, \label{eqn:FE0a}\\
		&\Delta^2\xi_{1,k}^{(0)}-\alpha_k\Delta\xi_{n,k}^{0}-\Wo^2\Delta\zeta_{1,k}^{0}=0. \label{eqn:FE0b}
	\end{align}
\end{subequations}
Before defining the boundary conditions on $\zeta_{1,k}^{(0)}$ and $\xi_{1,k}^{(0)}$ we first need to separate the $\widehat{F}_k$ coefficients, see \cref{eqn:BCendsFourZ}, into real and imaginary parts. The details for this separation can be found in \cref{sec:AppRealImag}. Then, $\zeta_{1,k}^{(0)}$ and $\xi_{1,k}^{(0)}$ must satisfy zero Neumann everywhere, zero Dirichlet on the cylinder, and $\zeta_{1,k}^{(0)} = \Re{\widehat{F}_k(y)}$ and $\xi_{1,k}^{(0)} = \Im{\widehat{F}_k(y)}$ on $x = \pm \frac{L}{2a}$ and on $y = \pm \frac{\ell}{2a}$.

As for pre-processing \cref{eqn:NewSS} we only need to simplify the nonlinear terms after expanding $\psi_{1,k}^{(0)}$ into real and imaginary components as $\zeta_{1,k}^{(0)}$ and $\xi_{1,k}^{(0)}$. Noting that these nonlinear terms in the sum are simply $2\Re{\nabla^\perp\psi_{1,j}^0\cdot\nabla\Delta\overline{\psi_{1,k-j}^0}}$ it is easy to see that they become $2\left(\nabla^\perp\zeta_{1,j}^{(0)}\cdot\nabla\Delta\zeta_{1,k-j}^{(0)}+\nabla^\perp\xi_{1,j}^{(0)}\cdot\nabla\Delta\xi_{1,k-j}^{(0)}\right).$

As for the truncation, we make use of the symmetry in $k$ to strictly consider $k \geq 0$. Therefore, after truncating and using the above expressions for the nonlinear terms in the sum, \cref{eqn:NewSS} becomes
\begin{equation}
	\Delta^2\psi_{0,k}^{(1)}-\alpha_k \Delta\psi_{0,k}^{(1)}=2\Wo^2\sum_{j=k-K}^K\left(\nabla^\perp\zeta_{1,j}^{(0)}\cdot\nabla\Delta\zeta_{1,k-j}^{(0)}+\nabla^\perp\xi_{1,j}^{(0)}\cdot\nabla\Delta\xi_{1,k-j}^{(0)}\right), \label{eqn:FESS}
\end{equation}
with zero Dirichlet and zero Neumann boundary conditions everywhere. 

For simplicity, we do not include the splitting of the higher order solutions $\psi_{2,k}^{(1)}$ and $\psi_{1,k}^{(2)}$ into real and imaginary parts as they are straightforward but cumbersome.


\subsection{Finite Element Method for the Biharmonic Equation}

Here we introduce the weak formulation for the $C^0$ penalty method we use for the biharmonic portion of \cref{eqn:FE0a,eqn:FE0b,eqn:FESS}. The formal derivation and rigorous analysis of this method can be found in \cite{engel2002continuous,babuvska1973nonconforming,brenner2005c}. Consider a domain $\Omega^* \in \R^2$ and the biharmonic equation $\Delta^2 \psi= f$ in $\Omega^*$ with $\psi = 0 = \partial_{\unitvect{n}} \psi$ on $\partial \Omega^*$. The primary idea behind this method is to look for $H^1$ solutions where only continuity between cells is required, but a penalty is enforced for discontinuity in the derivatives across cells. Start by defining the function space  $ V = \left\{ \phi \in H^1(\Omega^*) : \ \phi = 0 \text{ on }\partial \Omega^* \right\}.$ Then, the weak formulation is to find $\psi\in V$ such that $b(\psi,\phi) = L(\phi)$ for all $\phi \in V$, where $L(\phi) =\int_{\Omega^*} f\phi d\x$ and 
\begin{equation}
\begin{aligned}
b(\psi,\phi) &= \sum_{T\in \tau_n} \int_T \Delta \psi \Delta \phi ~ d\x \\
& \quad  - \sum_{\tilde{e}}  \left(\int_{\tilde{e}}  \Delta \psi \partial_{\unitvect{n}} \phi ~ ds + \int_{\tilde{e}} \Delta \phi\partial_{\unitvect{n}} \psi ~ ds - \frac{\alpha}{|\tilde{e}|}\int_{\tilde{e}} \partial_{\unitvect{n}} \psi \partial_{\unitvect{n}} \phi~ ds \right) \\
& \quad  - \sum_{\overline{e}}  \left(\int_{\overline{e}} \langle \Delta \psi \rangle \llbracket\nabla \phi \rrbracket ~ ds + \int_{\overline{e}} \langle \Delta \phi \rangle \llbracket\nabla \psi \rrbracket ~ ds - \frac{\alpha}{|\overline{e}|}\int_{\overline{e}} \llbracket \nabla \psi \rrbracket \llbracket \nabla \phi \rrbracket ~ ds \right).
\end{aligned}
\label{eqn:bihar_bilin}
\end{equation}
The triangularization of $\Omega^*$ is denoted by $\tau_n$ and the the interior edges and the exterior edges of each cell are denoted by $\overline{e}$ and $\tilde{e}$, respectively. In the integrals along the edges, $\llbracket \vect{v} \rrbracket = \left( \vect{v}_+ \cdot \unitvect{n} - \vect{v}_- \cdot \unitvect{n} \right)$ and $\langle f \rangle = \frac{1}{2} \left( f_+ + f_-\right)$ where the $+$ ($-$) subscript indicates that we are evaluating the function in the cell where the normal vector along that edge points into (out of) that cell. The penalty parameter $\alpha > 0$ was empirically chosen to be 20 for all results presented in this paper. Other values of $\alpha$ were tested, but appeared to have negligible impact on the solution near the cylinder.


\subsection{Weak Formulation}

We define the function spaces 
\begin{align*}
V_k^R &= \left\{ f \in H^1(\Omega) :\  f= \Re{\widehat{F}_k(y)} \text{ on }\Gamma_w \text{ and } f = 0 \text{ on }\Gamma_c \right\}, \\
V_k^I &= \left\{ f \in H^1(\Omega) :\  f= \Im{\widehat{F}_k(y)} \text{ on }\Gamma_w \text{ and } f = 0 \text{ on }\Gamma_c \right\}, \\
V_k &= V_k^R \times V_k^I,
\end{align*}
where $\Gamma_w$ represents the walls and $\Gamma_c$ represents the circle. Next, we let $u_k \in V_k^R$ and $v_k \in V_k^I$ be arbitrary test functions, multiply \cref{eqn:FE0a} by $u_k$ and \cref{eqn:FE0b} by $v_k$, and integrate over $\Omega$. After integrating by parts, we add the resulting equations. Then, the weak formulation for \cref{eqn:FE0a,eqn:FE0b} is to find $\left(\zeta_{1,k}^{(0)}, \xi_{1,k}^{(0)}\right) \in V_k$ such that 
\begin{equation*}
\begin{aligned}
b\left(\zeta_{1,k}^{(0)}, u_k\right) &- \alpha_k c\left(\zeta_{1,k}^{(0)},u_k\right) + \Wo^2 c\left(\xi_{1,k}^{(0)},u_k\right) \\ &+ b\left(\xi_{1,k}^{(0)},v_k\right) - \alpha_k c\left(\xi_{1,k}^{(0)},v_k\right) - \Wo^2 c\left(\zeta_{1,k}^{(0)},v_k\right) = 0
\end{aligned}
\end{equation*}
for all $(u_k, v_k) \in V_k$, where $b(\cdot, \cdot)$ is defined in \cref{eqn:bihar_bilin} and $c(\psi,\phi) = - \int_{\Omega} \nabla \psi \cdot \nabla \phi ~d\x$ is the bilinear form for the Laplacian components of the equation. 

For \cref{eqn:FESS} we define the function space $W_k = \left\{ f \in H^1(\Omega) :\  f=0 \text{ on } \partial \Omega \right\}$ then the weak formulation is to find  $\psi_{0,k}^{(1)} \in W_k$ such that 
\begin{equation*}
b\left(\psi_{0,k}^{(1)},w_k\right) - \alpha_k c\left(\psi_{0,k}^{(1)},w_k\right) = 2\Wo^2 \sum_{j = k-K}^K L_{j,k}(w_k),
\end{equation*}
for all $w_k \in W_k$, where 
\begin{equation*}
L_{j,k}(w_k) = \int_{\Omega} \left(\nabla^\perp\zeta_{1,j}^{(0)}\cdot\nabla\Delta\zeta_{1,k-j}^{(0)}+\nabla^\perp\xi_{1,j}^{(0)}\cdot\nabla\Delta\xi_{1,k-j}^{(0)} \right) w_k ~d\x.
\end{equation*}

These weak formulations are then solved via FEniCS \cite{LoggEtal102012}. 


\subsection{Convergence Results} \label{sec:NumericalConvergence}

There are two errors introduced numerically. The first is due to the truncation of the Fourier series in $z$, and the second is due to the finite element discretization error. 

\begin{figure}[tbhp]
\centering
\subfloat[Convergence in $z$-Fourier modes ]{\label{fig:FourConv} \includegraphics[width = 0.49\textwidth]{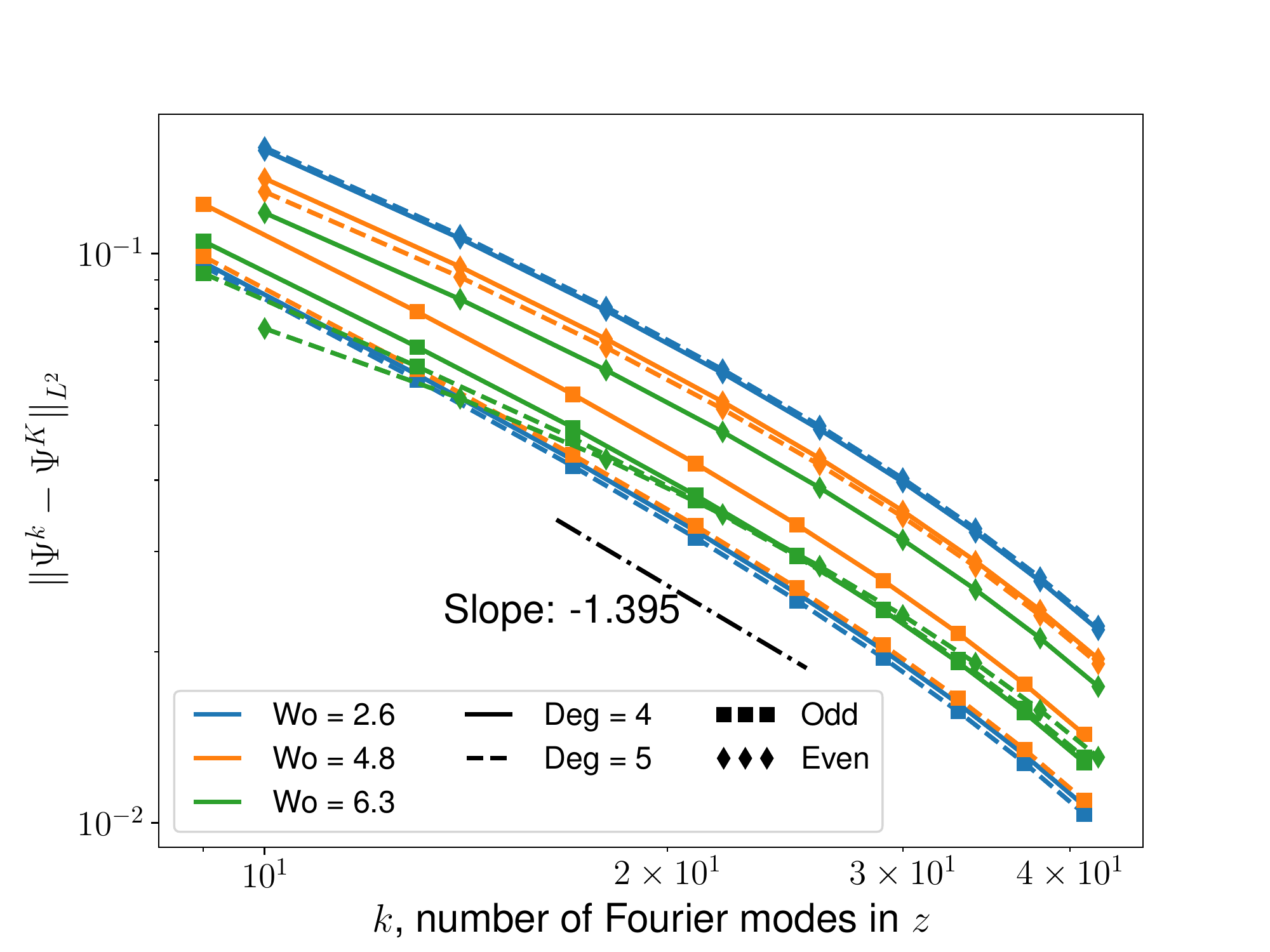}}
\subfloat[Mesh convergence]{\label{fig:MeshConv}\includegraphics[width = 0.49\textwidth]{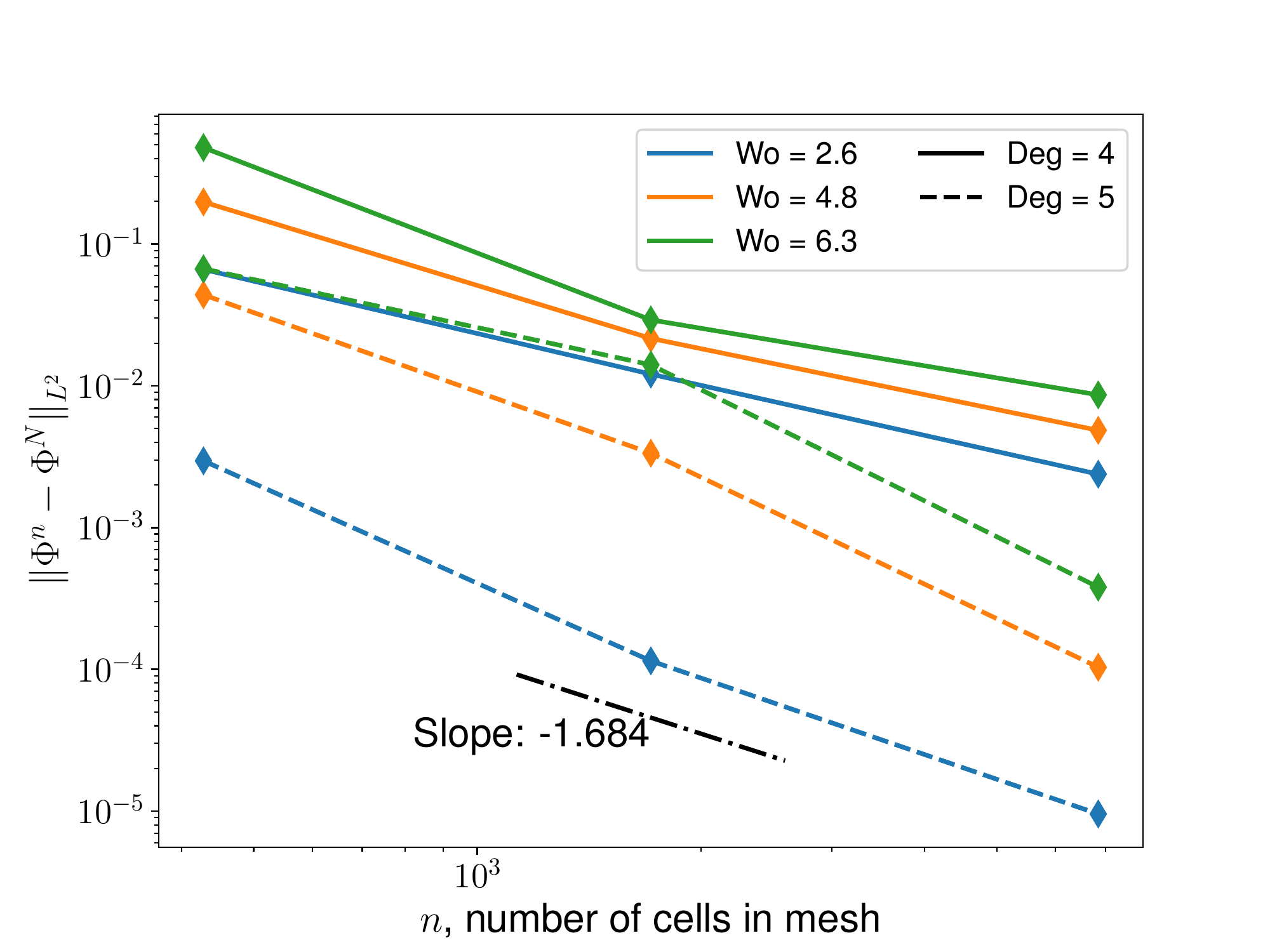}}
\caption{$L^2$ convergence of the steady streaming solution evaluated at the midheight of the channel ($z=0$), in (a) $z$-Fourier modes and (b) the number of cells in the mesh for $\Wo \approx 2.6,\ 4.8,\ 6.3$. The true solution is considered to be (a) $\Psi^K$ with $K=79$ for odd terms and  $K=80$ for even terms, and (b) $\Phi^N$ with $N=27392$. Both plots includes finite elements with polynomial degrees 4 and 5.}
\label{fig:Conv}
\end{figure}

Considering the error in the steady streaming solution arising from truncating the $z$-Fourier series, we define $\ds \Psi^K = \sum_{k = -K}^{K}  \psi_{0,k}^{(1)} e^{i\frac{4 \pi a}{h} k z}$ to be the truncated steady streaming stream function, such that $\Psi^K \to \psi_{0}^{(1)}$ as $K\to\infty$. To calculate the error we first evaluate at $z=0$ and then compute the error in the $L^2(\Omega)$-norm. The post-processed solutions are used for the convergence plots as these are the solutions presented in \cref{sec:results}. In \cref{fig:slipz_vsn2} at $z=0$ the post-processed solutions alternate back-and-forth across the unprocessed solution at $z=0$. For this reason, it is necessary to consider the convergence of $\Psi^K$ separately for even $K$ versus odd $K$. The true solution is therefore considered to be $\Psi^K$ with $K = 79$ for odd $k$ and $K=80$ for even $k$ and the convergence plot for $\Psi^k$ is shown in \cref{fig:FourConv}. We observe a rate of convergence of approximately 1.4 in the $z$-Fourier modes. For all results in \cref{sec:results}, we consider $K=10$. 

For the discretization error, we define $\ds \Phi^n = \sum_{k = -K}^{K}  \psi_{0,k}^{(1),n} e^{i\frac{2 \pi a}{h} k z},$ where the superscript $n$ denotes the number of cells in the mesh, which can be thought of as a proxy for the dimension of the discretized functional space. As $n$ is increased, the mesh is refined hierarchically using the FEniCS function \texttt{refine}, which retains the current vertices and adds new vertices to split all cells into four subcells. The refined mesh's minimum cell diameter and maximum cell diameter are exactly half of those values for the original mesh. The convergence plot for $\Phi^n$ is shown in \cref{fig:MeshConv}, where the true solution is considered to be $\Phi_N$ with $N=27392$ and $K=10$. Overall, we observe a rate of convergence of approximately 1.7 as we increase the number of cells in the mesh.

We consider both degree 4 and degree 5 polynomial spaces for the finite elements in both numerical convergence tests illustrated in \cref{fig:Conv}. For the convergence in the $z$-Fourier modes, \cref{fig:FourConv}, we see that the polynomial degree had a negligible effect here, as expected. For the convergence in the number of cells in the mesh, \cref{fig:MeshConv}, there is a noticeable difference between the results when using degree 4 polynomials versus using degree 5 polynomials. However, the rate of convergence for both is faster than first-order (approximately 1.57 for degree 4 polynomials and 1.80 for degree 5 polynomials). For all results in \cref{sec:results}, we consider degree 4 polynomials, with the exception of the higher-order solutions which will be discussed in \cref{sec:results}. 

\section{Results} \label{sec:results} 

Here we will use our model to analyze the steady streaming solution. First, we will discuss how the steady streaming solution depends on the shape of the domain in the $x$ and $y$ directions. We will also evaluate how the steady streaming solution depends on both the $z$-height within the channel and the frequency. Finally, we will conclude the results by considering the higher-order solutions. Unless otherwise specified, e.g., \cref{fig:tangVel_znorm,fig:TangVel_z}, all solutions are evaluated at the midheight of the channel, $z=0$.

\begin{figure}[tbhp]
    \centering
    \subfloat[$\Wo \approx 2.6$]{\label{fig:AllQuad2_6} \includegraphics[width = 0.49\textwidth,trim = {4cm 0.75cm 0.50cm 1.5cm},clip=true]{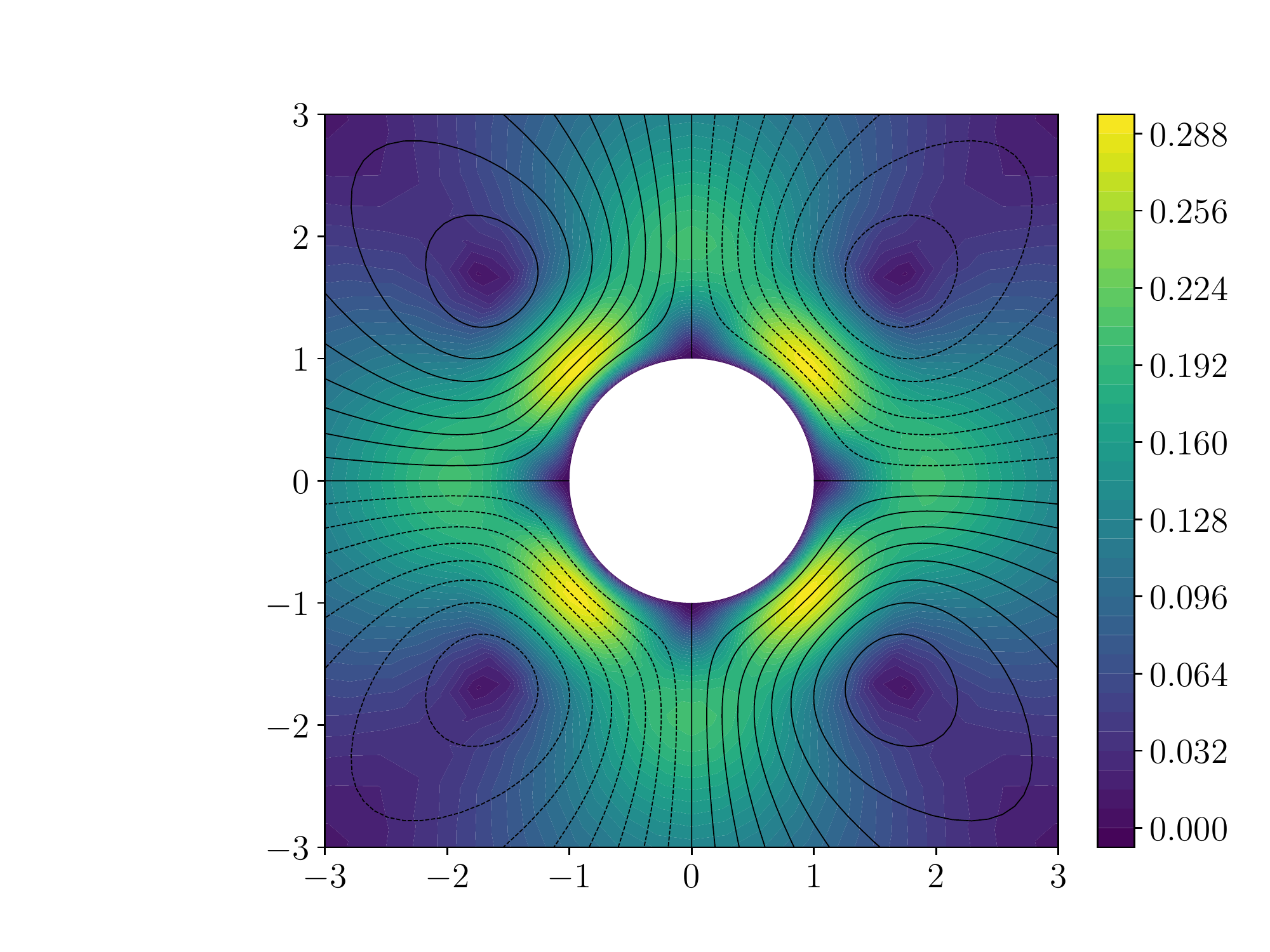}}
    \subfloat[$\Wo \approx 6.3$]{\label{fig:AllQuad6_3} \includegraphics[width = 0.49\textwidth,trim = {4cm 0.75cm 0.50cm 1.5cm},clip=true]{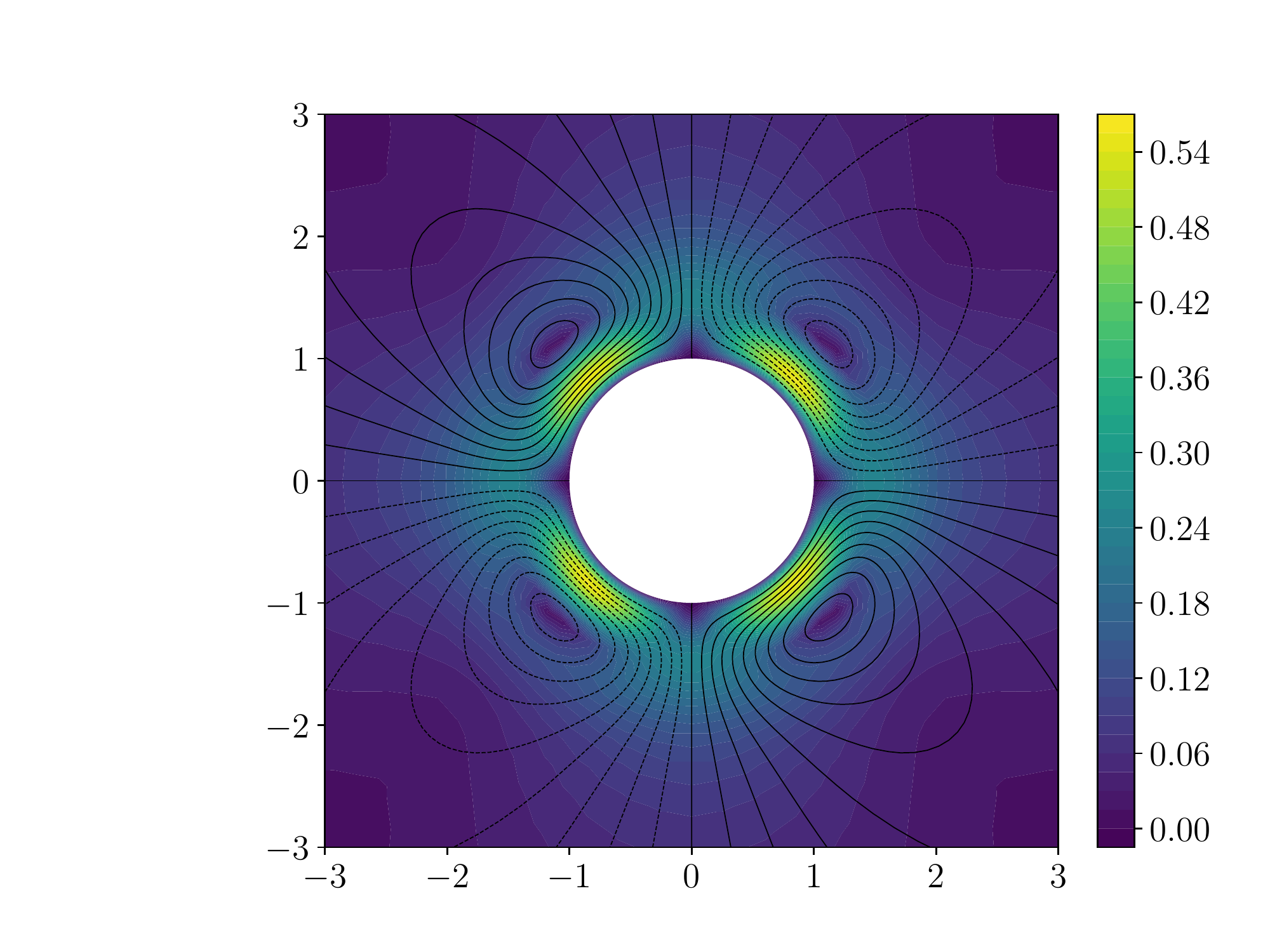}}
    \caption{Steady streaming velocity magnitude for $\Wo \approx 2.6,\ 6.3$ plotted in all four quadrants to demonstrate four-fold symmetry and vortex direction.}
    \label{fig:AllQuad}
\end{figure}

In \cref{fig:AllQuad,fig:WallTestFixedRatio}, we plot the steady streaming velocity magnitude with the streamlines superimposed on top. In terms of the stream function, the velocity magnitude is $\sqrt{\left(\partial_x \psi \right)^2 + \left(\partial_y \psi\right)^2 }$ and the streamlines are the contours of the stream function. In \cref{fig:AllQuad}, we observe negatively oriented vortices, i.e., clockwise, in quadrants 1 and 3, denoted by dashed lines. In contrast, we have positively oriented vortices, i.e., counter-clockwise, in quadrants 2 and 4, denoted by solid lines. The direction of the vortices is consistent with the experiment results on Newtonian fluids in \cite{vishwanathan2019steady,vishwanathan2019steadyViscometry}. Finally, we note the symmetry by a rotation of $\frac{\pi}{2}$ for both the velocity magnitude and streamlines in \cref{fig:AllQuad}. We will refer this symmetry as {\it four-fold symmetry} for the remainder of this discussion. 

A breaking of this four-fold symmetry has been recently observed experimentally \cite{vishwanathan2019steady,vishwanathan2019steadyViscometry}. In these experiments, an area of high velocity is observed north and south of the cylinder, which is not present east and west of the cylinder. To be explicit, we take north as the positive $y$ direction and east as the positive $x$ direction. Further, by area of high velocity, we mean that the velocity there is near to the maximum velocity or is the maximum velocity. It was suggested that this interesting observation \cite{vishwanathan2019steady,vishwanathan2019steadyViscometry} was due to the fact the walls are nearer to the cylinder in the $y$ direction than they are in the $x$ direction. We use our method to test this idea in \cref{sec:DomainDependence} and generally analyze how the steady streaming near the cylinder changes as the domain width changes.

\subsection{Domain Dependence }\label{sec:DomainDependence}

\begin{figure}[tbhp]
\centering
	\subfloat[\\$\Wo \approx 2.6$, $\frac{\ell}{a} = 50$]{\label{fig:WallTestRatio_Wo2_6a}
	\includegraphics[width = 0.32\textwidth,trim = {4cm 0.75cm 0.50cm 1.5cm},clip=true]{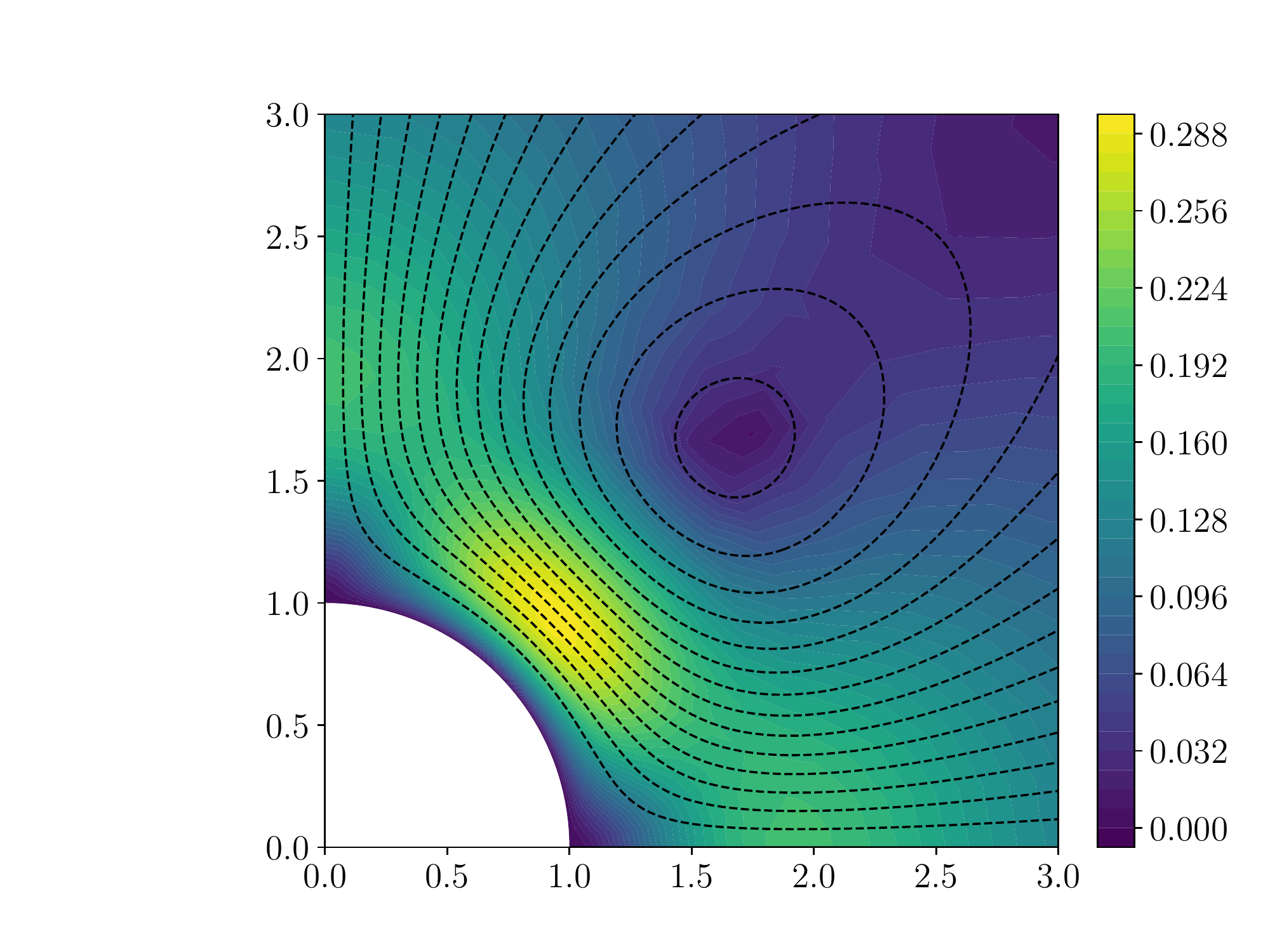}}
	\subfloat[\\$\Wo \approx 2.6$, $\frac{\ell}{a} = 10$]{\label{fig:WallTestRatio_Wo2_6b}
	\includegraphics[width = 0.32\textwidth,trim = {4cm 0.75cm 0.50cm 1.5cm},clip=true]{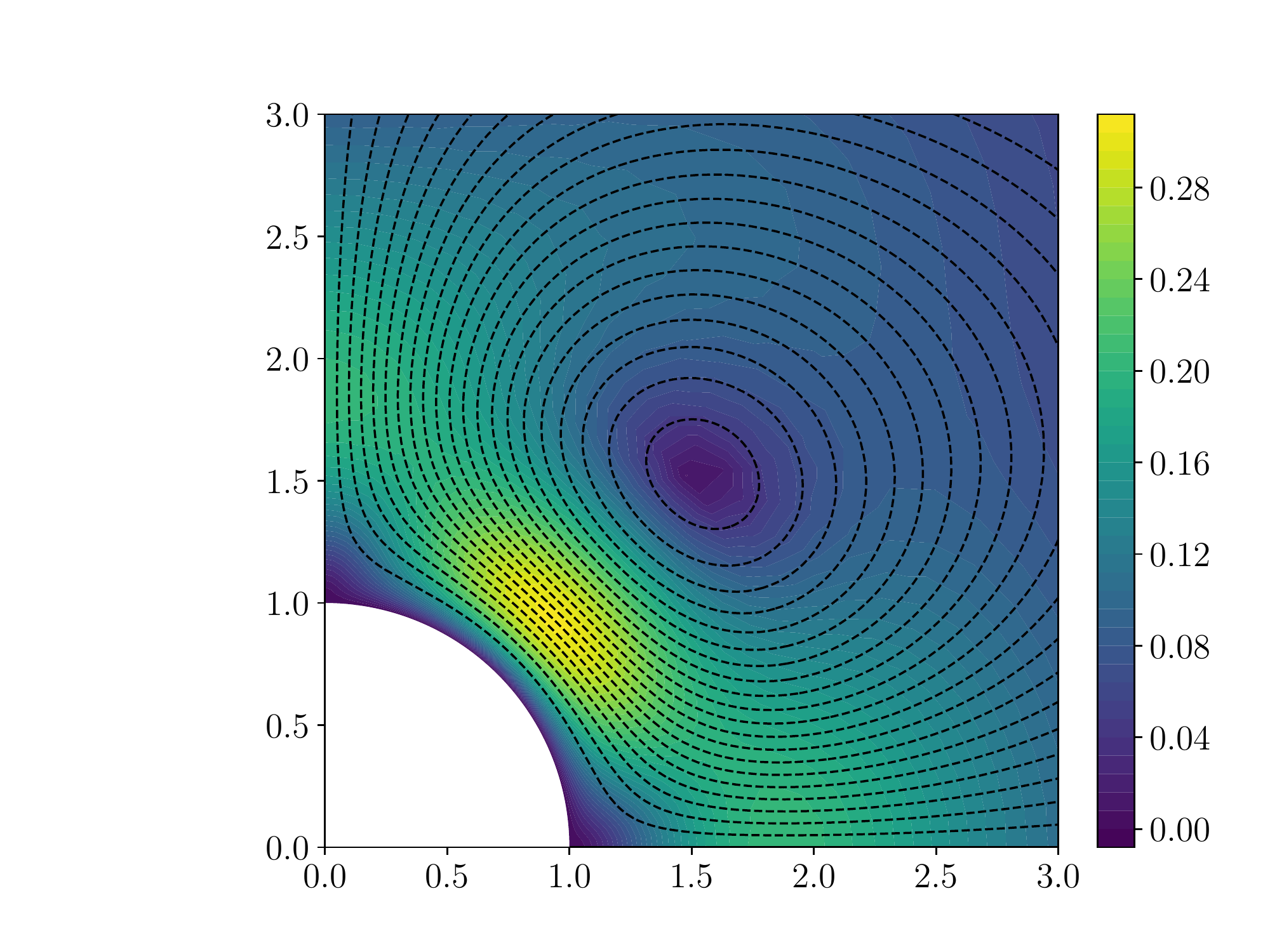}}
	\subfloat[\\$\Wo \approx 2.6$, $\frac{\ell}{a} = 4$]{\label{fig:WallTestRatio_Wo2_6c}
	\includegraphics[width = 0.32\textwidth,trim = {4cm 0.75cm 0.50cm 1.5cm},clip=true]{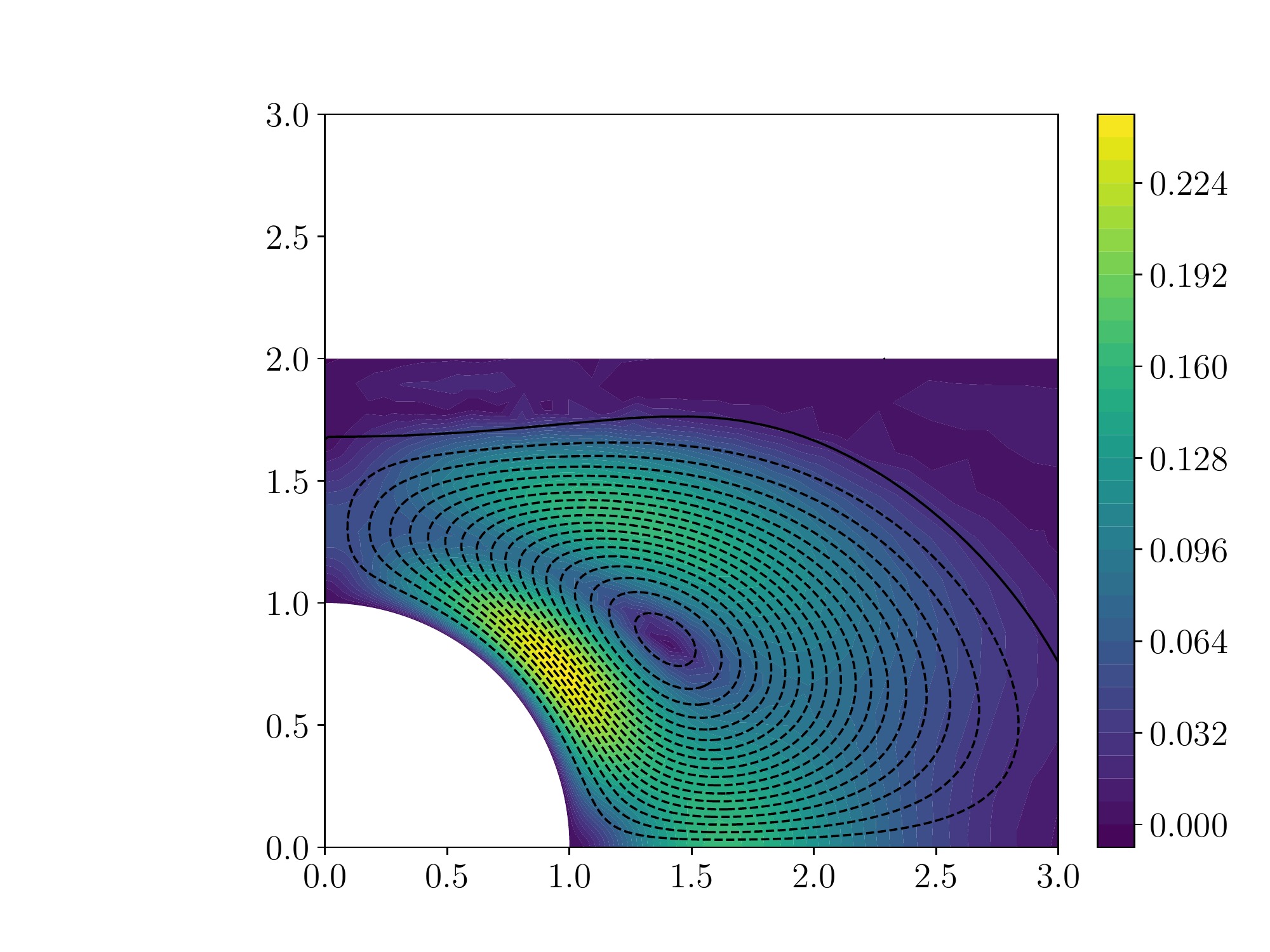}}

 	\subfloat[\\$\Wo \approx 4.8$, $\frac{\ell}{a} = 50$]{\label{fig:WallTestRatio_Wo4_8a}
	\includegraphics[width = 0.32\textwidth,trim = {4cm 0.75cm 0.50cm 1.5cm},clip=true]{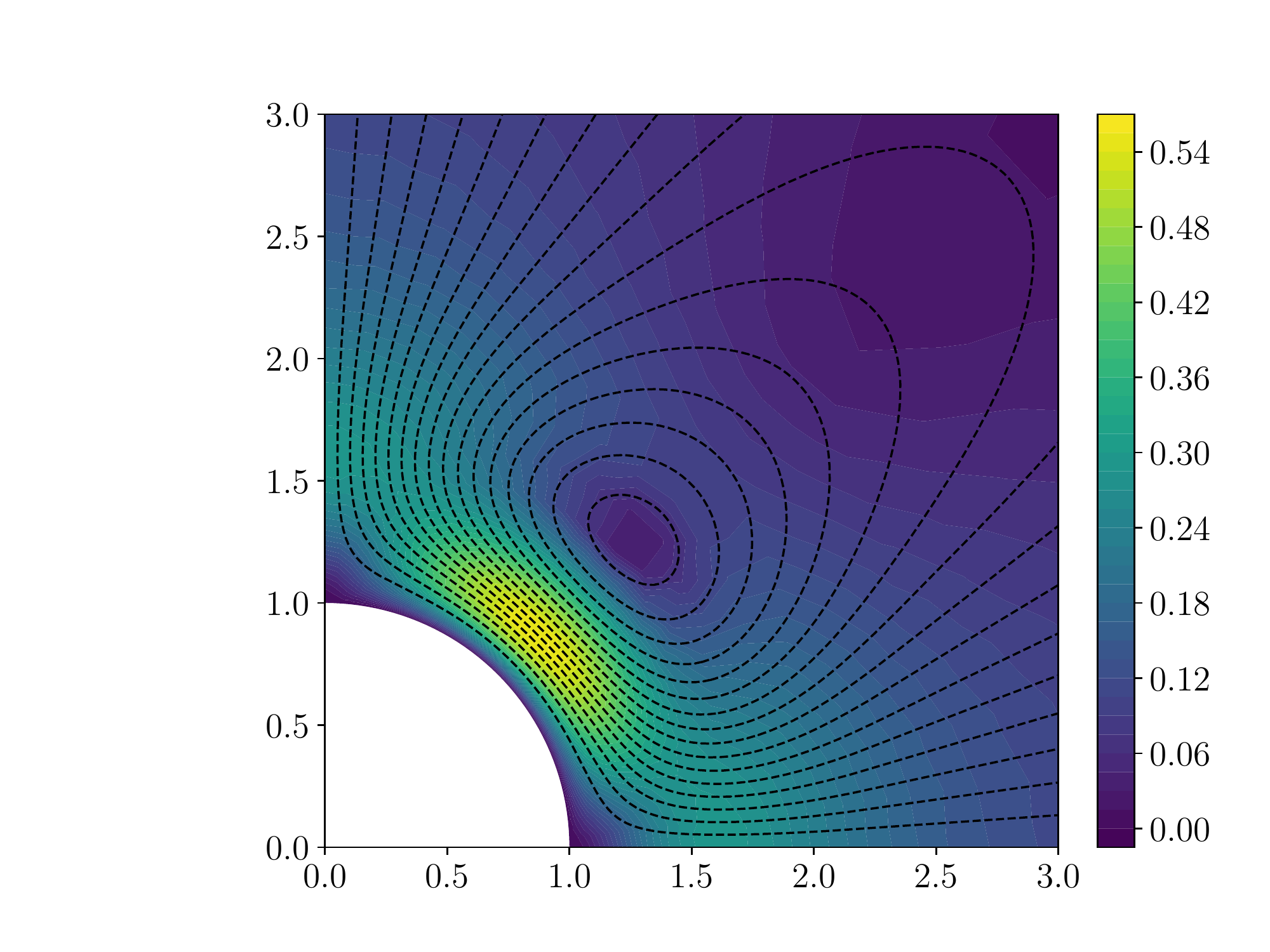}}
	\subfloat[\\$\Wo \approx 4.8$, $\frac{\ell}{a} = 10$]{\label{fig:WallTestRatio_Wo4_8b}
	\includegraphics[width = 0.32\textwidth,trim = {4cm 0.75cm 0.50cm 1.5cm},clip=true]{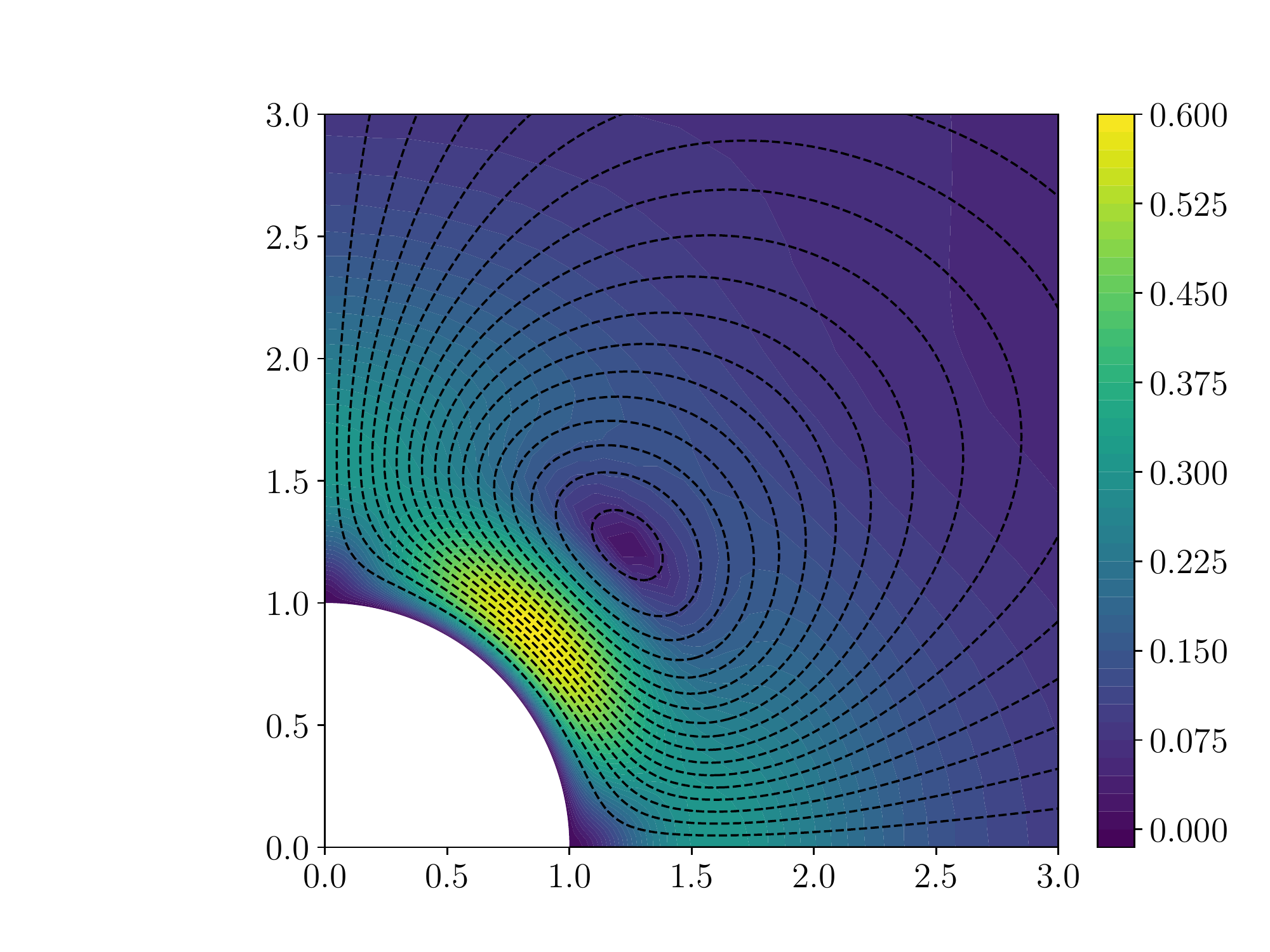}}
	\subfloat[\\$\Wo \approx 4.8$, $\frac{\ell}{a} = 4$]{\label{fig:WallTestRatio_Wo4_8c}
	\includegraphics[width = 0.32\textwidth,trim = {4cm 0.75cm 0.50cm 1.5cm},clip=true]{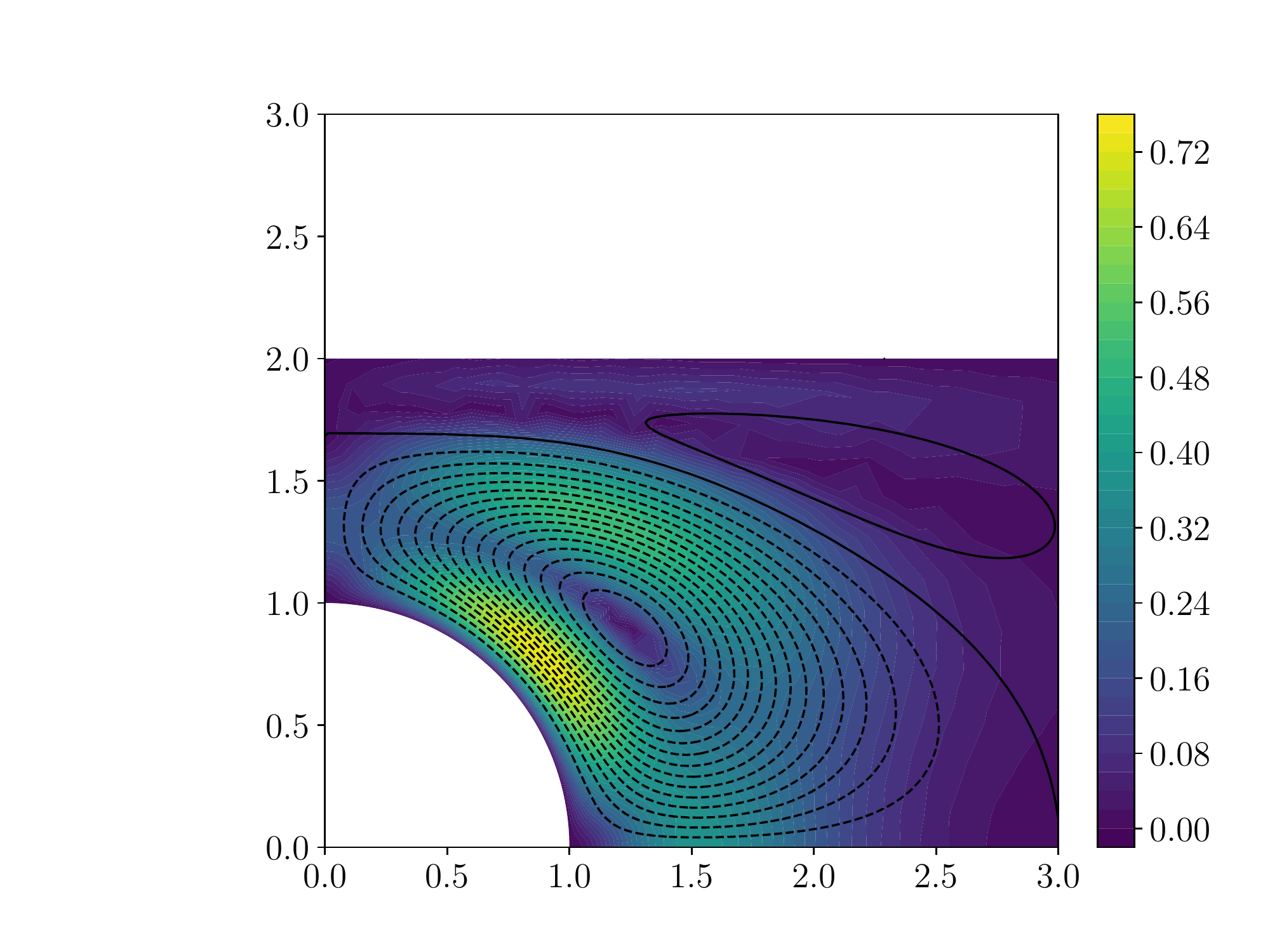}}

 	\subfloat[\\$\Wo \approx 6.3$, $\frac{\ell}{a} = 50$]{\label{fig:WallTestRatio_Wo6_3a}
	\includegraphics[width = 0.32\textwidth,trim = {4cm 0.75cm 0.50cm 1.5cm},clip=true]{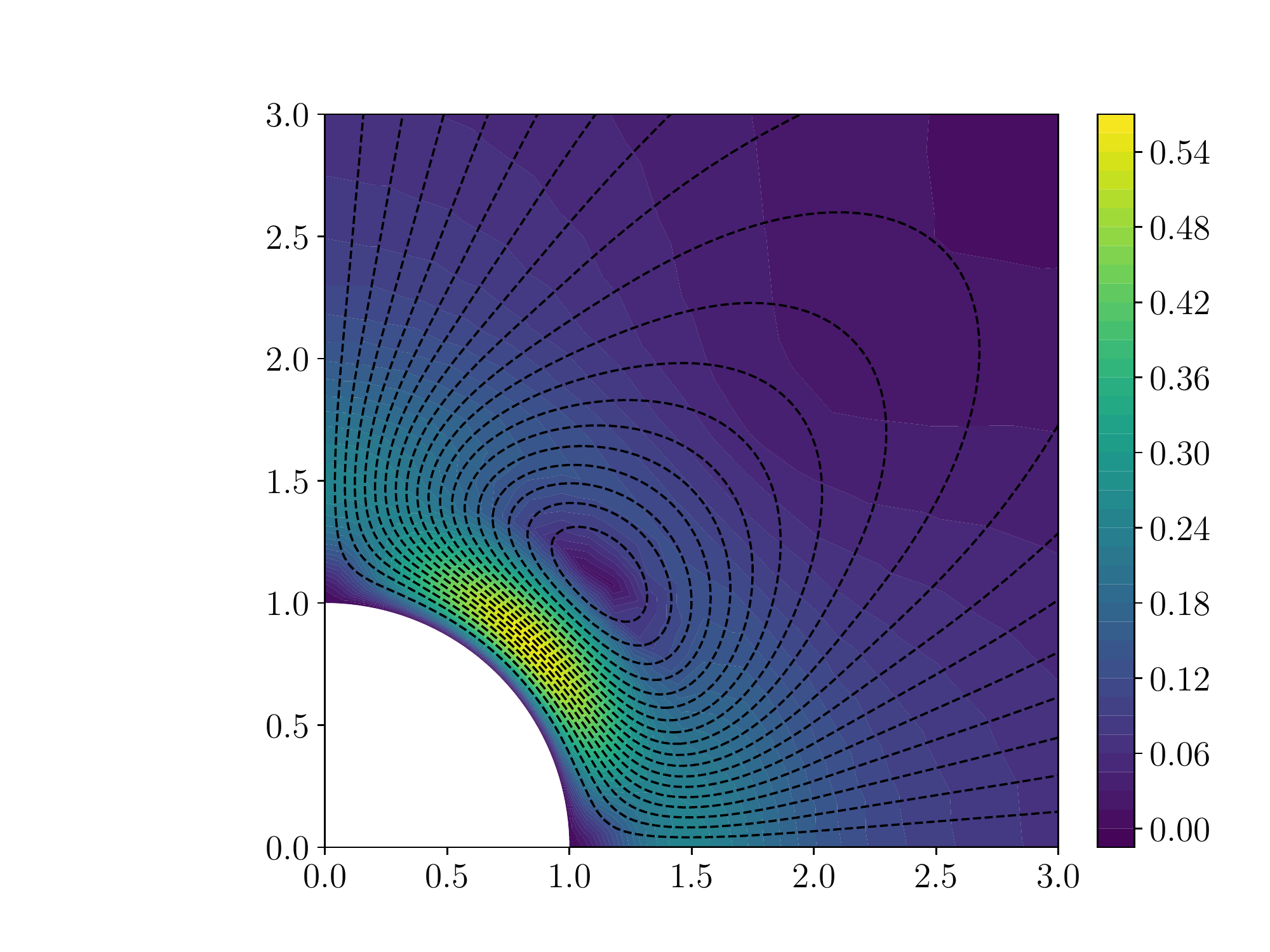}}
	\subfloat[\\$\Wo \approx 6.3$, $\frac{\ell}{a} = 10$]{\label{fig:WallTestRatio_Wo6_3b}
	\includegraphics[width = 0.32\textwidth,trim = {4cm 0.75cm 0.50cm 1.5cm},clip=true]{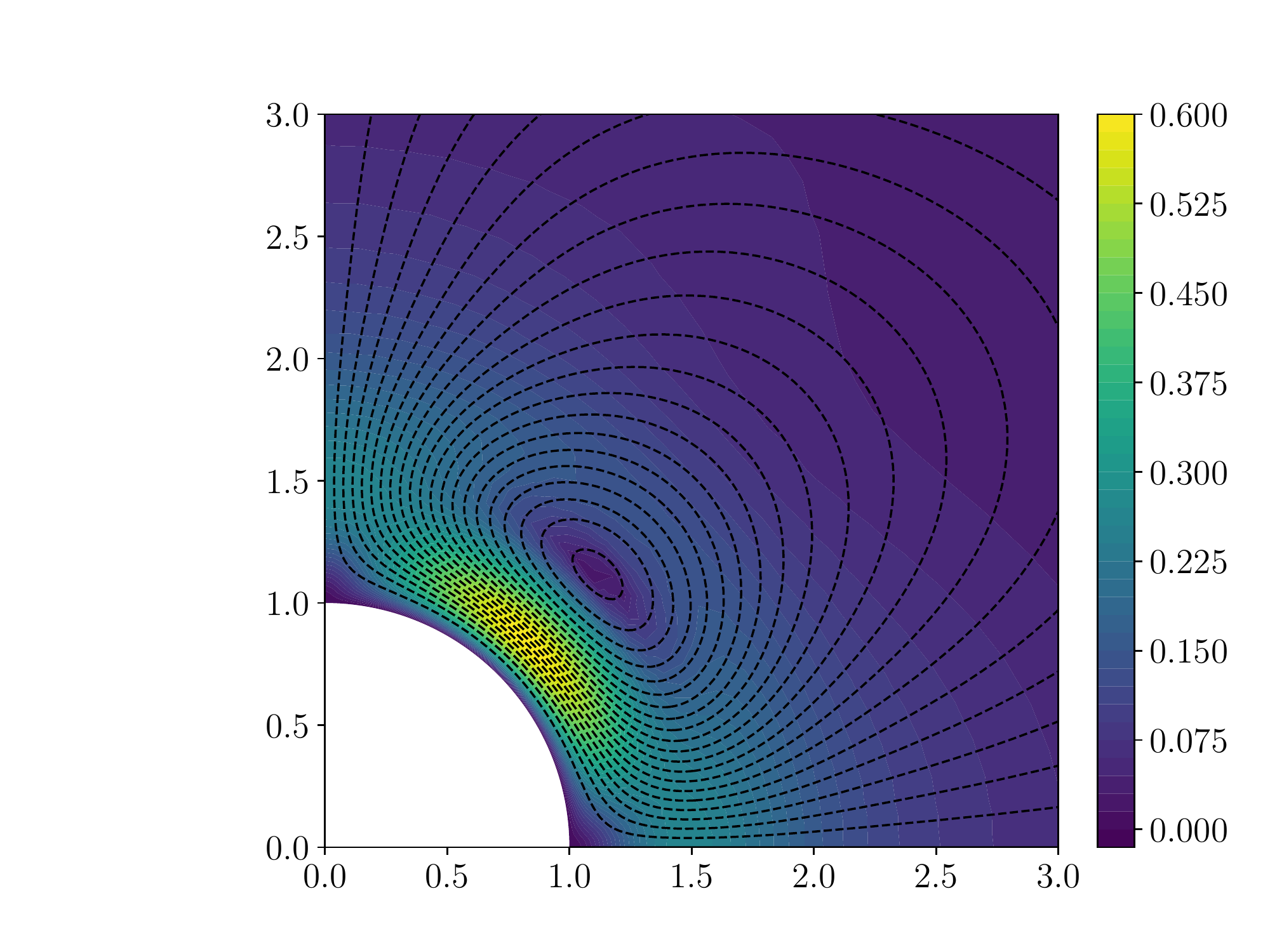}}
	\subfloat[\\$\Wo \approx 6.3$, $\frac{\ell}{a} = 4$]{\label{fig:WallTestRatio_Wo6_3c}
	\includegraphics[width = 0.32\textwidth,trim = {4cm 0.75cm 0.50cm 1.5cm},clip=true]{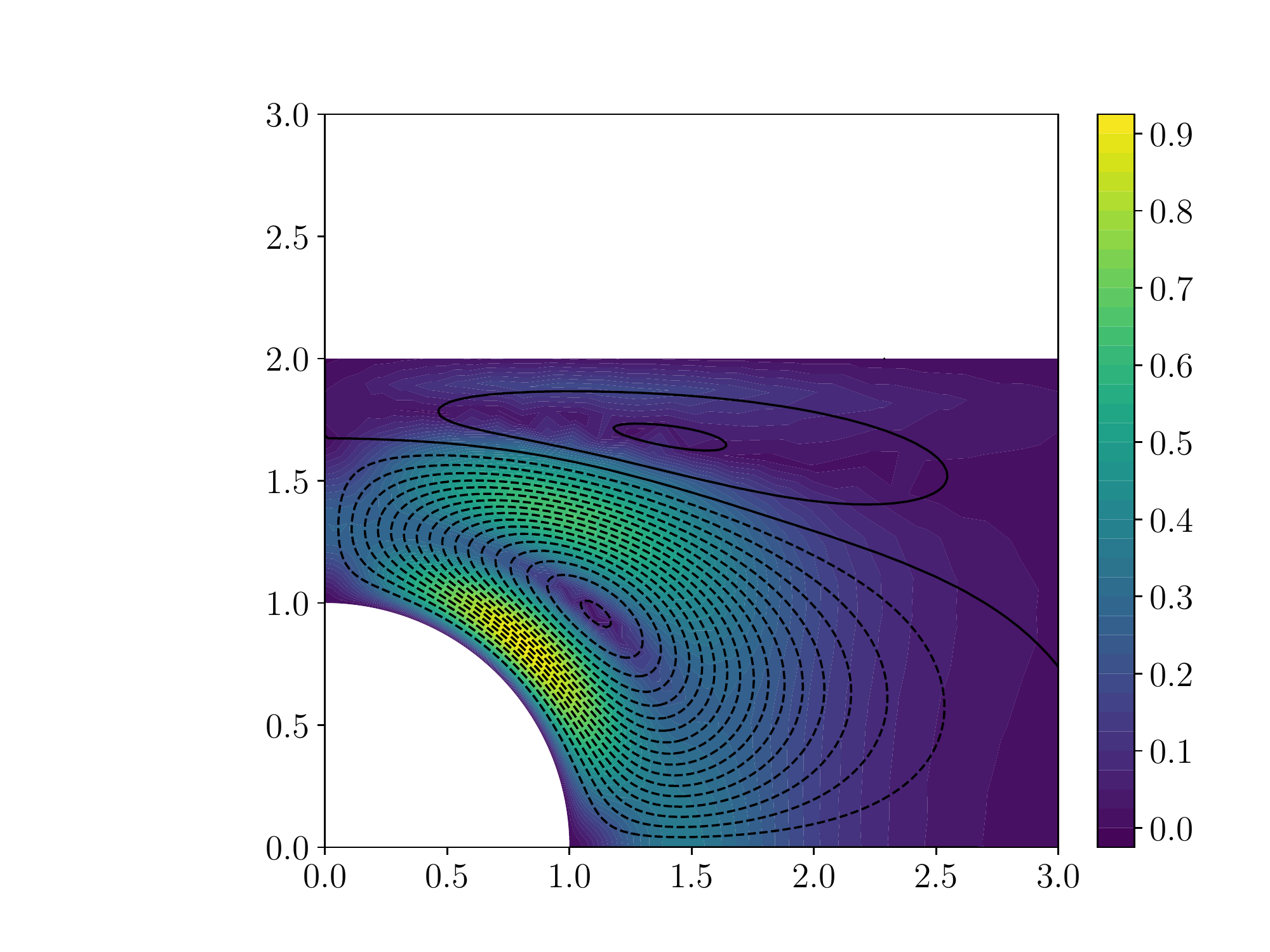}}
\caption{Steady streaming velocity magnitude for $\Wo \approx 2.6, 4.8, 6.3$ (rows) and $\frac{\ell}{a} = 50,10,4$ (columns) while keeping $\frac{L}{\ell}$ fixed at 4.}
\label{fig:WallTestFixedRatio}
\end{figure}

In \cref{fig:WallTestFixedRatio} we plot the steady streaming velocity as we make the channel more narrow in the $y$-direction, i.e., $\ell$ transitions form $\ell \gg a$ to $\ell = O(a)$. The $\Wo$ is varied across the rows of \cref{fig:WallTestFixedRatio}, i.e., the $\Wo$ is increased going from \cref{fig:WallTestRatio_Wo2_6a} to \cref{fig:WallTestRatio_Wo4_8a} to \cref{fig:WallTestRatio_Wo6_3a}, and the width of the domain is varied across the columns \cref{fig:WallTestFixedRatio}, i.e., $\ell$ is decreased going from \cref{fig:WallTestRatio_Wo2_6a} to \cref{fig:WallTestRatio_Wo2_6b} to \cref{fig:WallTestRatio_Wo2_6c}. Two cases were considered when changing the domain shape. The first was to decrease $\ell$ while keeping the aspect ratio fixed at $\frac{L}{\ell}=4$ and the second was to keep $L$ constant such that $\frac{L}{a}=200$ was fixed. The differences between the $\frac{L}{\ell} = 4$ and $\frac{L}{a} = 200$ results were negligible so we only include the fixed aspect ratio test, which are the results shown in \cref{fig:WallTestFixedRatio}. 

As for the breaking of the four-fold symmetry, we first note that all of the domains are rectangular with $L>\ell$, so none of the solutions truly have this four-fold symmetry throughout the entire domain, but if we restrict this analysis to be near the cylinder then this symmetry is observed. For domains where the $x$ and $y$ walls are sufficiently far away from the cylinder, i.e., $\frac{L}{a},\ \frac{\ell}{a} \gg 1$ as in \cref{fig:AllQuad}, the solutions have the four-fold symmetry. The solutions in \cref{fig:WallTestRatio_Wo4_8a,fig:WallTestRatio_Wo2_6b,fig:WallTestRatio_Wo4_8b,fig:WallTestRatio_Wo6_3b} also demonstrate the four-fold symmetry when all four quadrants are plotted, but we restrict ourselves to the first quadrant to better illustrate the vortex structure. To investigate any possible breaking of the four-fold symmetry we focus on velocity north and south of the cylinder as compared to the velocity east and west of the cylinder. However, in none of the scenarios in \cref{fig:WallTestFixedRatio} do we observe velocities on the $y$ axis being noticeably larger than velocities on the $x$ axis, even when the walls are only a single radii away from the cylinder. Further, it appears that in those extreme cases of $\frac{\ell}{a}=4$ the velocity is actually larger on the $x$ axis than on the $y$ axis. Therefore, we conclude that, at least through our method, the areas of high-velocity north and south of the cylinder, and the corresponding symmetry breaking, observed in \cite{vishwanathan2019steady,vishwanathan2019steadyViscometry} are not due to the shape of the domain. However, the true reason for these observations remains open. We want to end this discussion by noting that with the exception of this unobserved area of high velocity, our solutions appear to agree reasonably well with the experimental results presented in \cite{vishwanathan2019steady,vishwanathan2019steadyViscometry}.

\begin{figure}[tbhp]
\centering 
\includegraphics[width = 0.4\textwidth,trim = {0.5cm 0.25cm 2cm 1.5cm},clip=true]{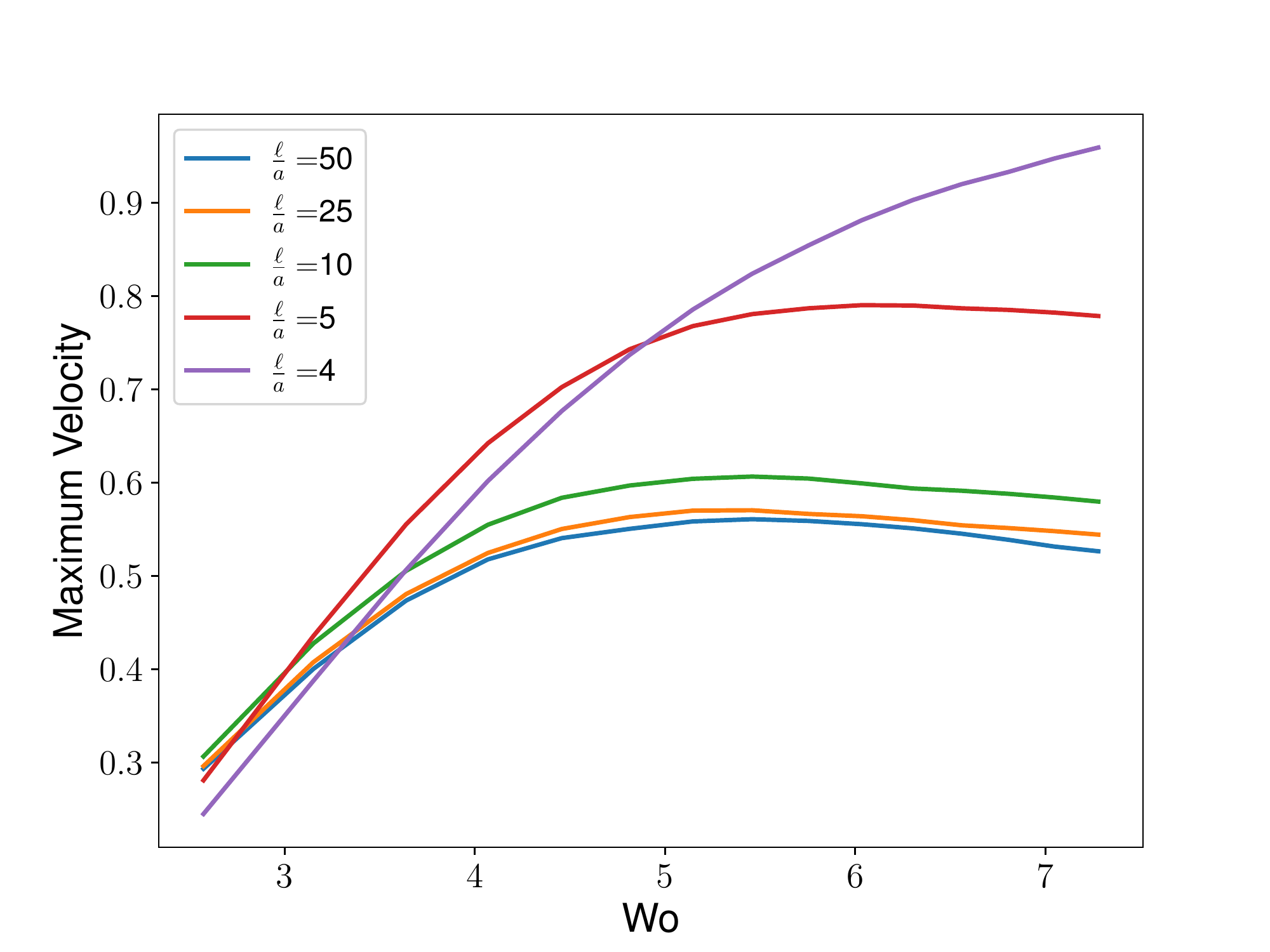}
\caption{Maximum steady streaming velocity versus the $\Wo$ for $\frac{\ell}{a} = 50,25,10,5,4$}
\label{fig:MaxVelDomain}
\end{figure}

We can see that as $\ell \to 4a$ the steady streaming solution retains the structure of four counter-rotating vortices near the cylinder. As $\frac{\ell}{a}$ is changed from $50 \rightarrow 10$, the maximum velocity magnitude for each $\Wo$ increases and the streamlines show the vortices becoming more circular, particularly in \cref{fig:WallTestRatio_Wo2_6b}, but other than that the solutions appear relatively unchanged in this scenario. In comparison, changing $\frac{\ell}{a}$ from $10 \rightarrow 4$ gives rise to a significant change in the solutions, as expected. It is unsurprising that the shape of the vortices is impacted with $\frac{\ell}{a} = 4$ as the walls are now one radii away from the cylinder, but it is worth noting how the vortices shrink faster in the $y$-direction than in the $x$-direction resulting in this elongated appearance. There is also a qualitative change in the velocity magnitude as $\frac{\ell}{a}$ changes from $10 \rightarrow 4$ between $\Wo = 2.6$ and $\Wo = 4.8$ and $6.3$, where it is observed that the lower $\Wo$ results in a decreased maximum velocity versus the higher $\Wo$ values resulting in an increased maximum velocity for the decreased domain size. Notably, comparing \cref{fig:WallTestRatio_Wo6_3b,fig:WallTestRatio_Wo6_3c} we see that the maximum velocity increases approximately 50\% as $\frac{\ell}{a}$ is decreased from $10 \rightarrow 4$. In \cref{fig:MaxVelDomain} the maximum steady streaming velocity versus the $\Wo$ is plotted for $\frac{\ell}{a} = 50,25,10,5,4$. For every test scenario completed, the maximum velocity (in each quadrant) was always realized between the cylinder and the vortex center, as can be seen in \cref{fig:AllQuad,fig:WallTestFixedRatio}. For each dimensionless channel width ($\frac{\ell}{a}$) tested the maximum velocity initially increases, but eventually, with the exception of $\frac{\ell}{a}=4$, the maximum velocity slowly decreases. The maximum velocity for $\frac{\ell}{a} = 4$ is monotonically increasing for the range of $\Wo$ tested, but it appears that for larger $\Wo$ this will either decrease as the other curves do or possibly asymptote. The solutions are qualitatively very similar for $\frac{\ell}{a}>10$, whereas there is a stark distinction for $\frac{\ell}{a}<10$. 

\begin{figure}[tbhp]
\centering 
\includegraphics[width = 0.99\textwidth,trim = {0cm 0.75cm 0cm 0.65cm},clip=true]{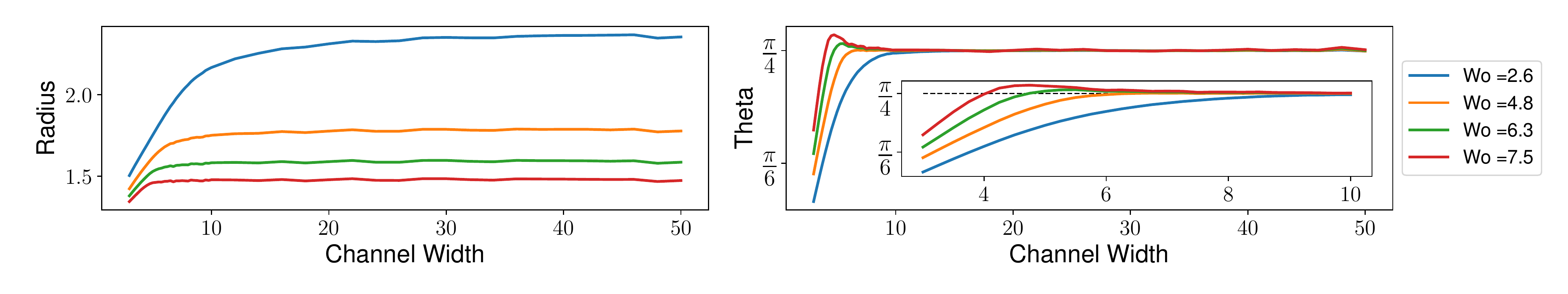}
\caption{Location, in polar coordinates, of the vortex center versus dimensionless channel width $\frac{l}{a}$ for $\Wo \approx 2.6, 4.8, 6.3, 7.5$. The inset plot on the theta plot (right) zooms in on the region where $\frac{l}{a} \in[3,10]$. The black dashed line at $\frac{\pi}{4}$ is included in the inset plot to clearly demonstrate that vortex center is at $\frac{\pi}{4}$ for all $\Wo$ when the walls are sufficiently far away form the cylinder.}
\label{fig:vortex}
\end{figure}

For a final analysis on how steady streaming depends on the domain shape we follow the vortex center as $\frac{\ell}{a}$ is decreased from $50 \rightarrow 3$ in \cref{fig:vortex}. Again, as $\ell$ is decreased we keep the aspect ratio $\frac{L}{\ell}=4$ fixed. We use the \texttt{optimize} library from SciPy \cite{2020SciPy-NMeth} to compute the center of the vortex which is the minimum of the stream function within the vortex. Taking advantage of the symmetry of the solutions we restrict the minimization process to the first quadrant. The location, in polar coordinates, of the vortex center versus the dimensionless channel width is shown in \cref{fig:vortex} with the radius on the left and the azimuth on the right. The vortices all move monotonically away from the cylinder monotonically as the outer walls are moved out as can be seen by the radius in \cref{fig:vortex} (left). The azimuth initially increases but is not monotonically increasing for $\Wo = 6.3, 7.5$; however for each $\Wo$ the azimuth asymptotes to $\frac{\pi}{4}$ around $\frac{l}{a} = 10$. The radii also converge to a constant at or before $\frac{l}{a}=10$ for $\Wo = 4.8, 6.3, 7.5$, but does not converge until a much larger value, $\frac{l}{a} \approx 35$, for $\Wo = 2.6$. This all suggests that if the walls are sufficiently far away ($\frac{l}{a} >10$) from the cylinder then the walls do not impact the steady streaming vortices.

For the remainder of \cref{sec:results} the domain will be $\frac{L}{a} = 200$ and $\frac{\ell}{a} =50$, which is analogous to the domain in \cite{vishwanathan2019steady}.

\subsection{Tangential Steady Streaming Velocity}
A standard \cite{vlassopoulos1993characterization,vishwanathan2019steady,chang1979secondary} method to analyze steady streaming is to evaluate the steady streaming solution along a radial line at $\theta = \frac{\pi}{4}$. The choice of the $\theta = \frac{\pi}{4}$ radial line is due to the fact that this line passes through the vortex center and all the streamlines are orthogonal to this line, i.e., all the flow is in the azimuthal direction. Therefore, the speed and direction of the flow can be characterized by a single scalar function along this line, which is simply the azimuthal component of the velocity vector in polar coordinates. In cartesian, we dot the velocity vectors along this line with $(-1,1)$, which is orthogonal to the radial line. As can be seen in \cref{fig:vortex} the azimuth of the vortex center asymptotes to $\frac{\pi}{4}$ for $\frac{\ell}{a}>10$ for all $\Wo$ presented, suggesting this is a robust method for this analysis. Following \cite{vishwanathan2019steady} we will refer to this process as the tangential steady streaming velocity.

\begin{figure}[tbhp]
\centering
	\subfloat[$\Wo \approx 2.6$]{\label{fig:TangVel_z_Wo2_6}
	\includegraphics[width = 0.24\textwidth,trim = {0.5cm 0.5cm 0.5cm 1.4cm},clip=true]{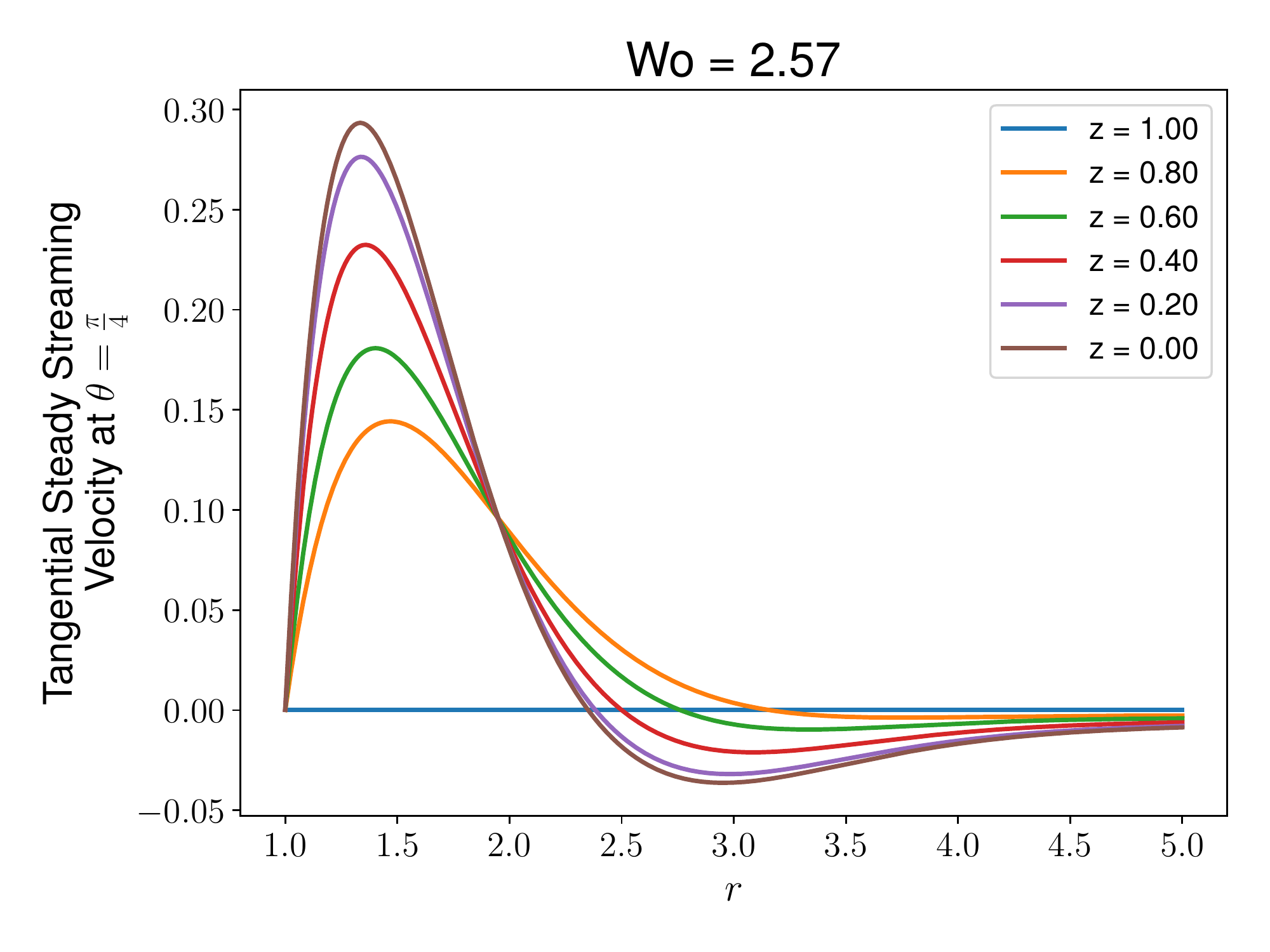}}
	\subfloat[$\Wo \approx 4.8$]{\label{fig:TangVel_z_Wo4_8}
	\includegraphics[width = 0.24\textwidth,trim = {0.5cm 0.5cm 0.5cm 1.4cm},clip=true]{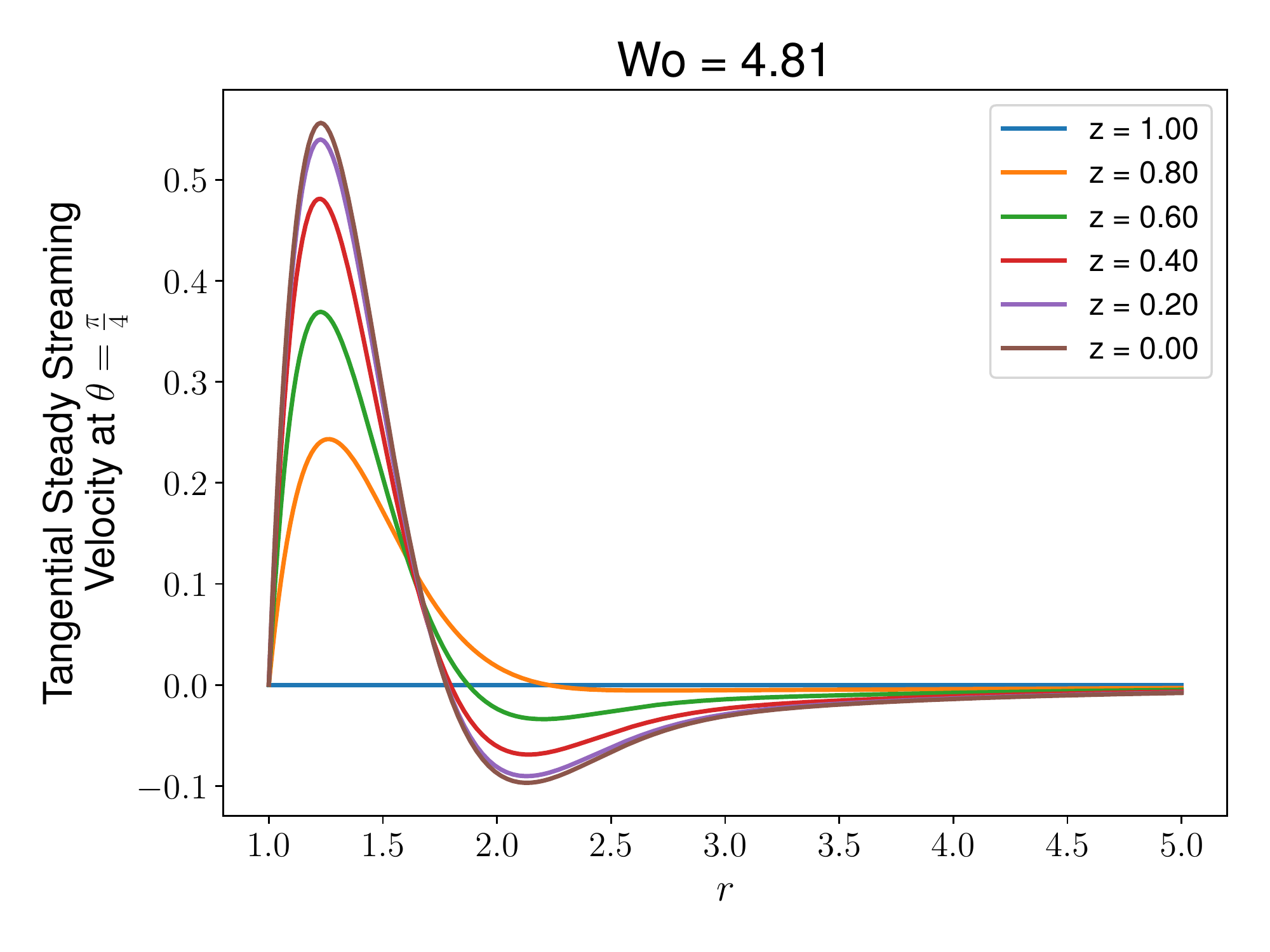}}
	\subfloat[$\Wo \approx 6.3$]{\label{fig:TangVel_z_Wo6_3}
	\includegraphics[width = 0.24\textwidth,trim = {0.5cm 0.5cm 0.5cm 1.4cm},clip=true]{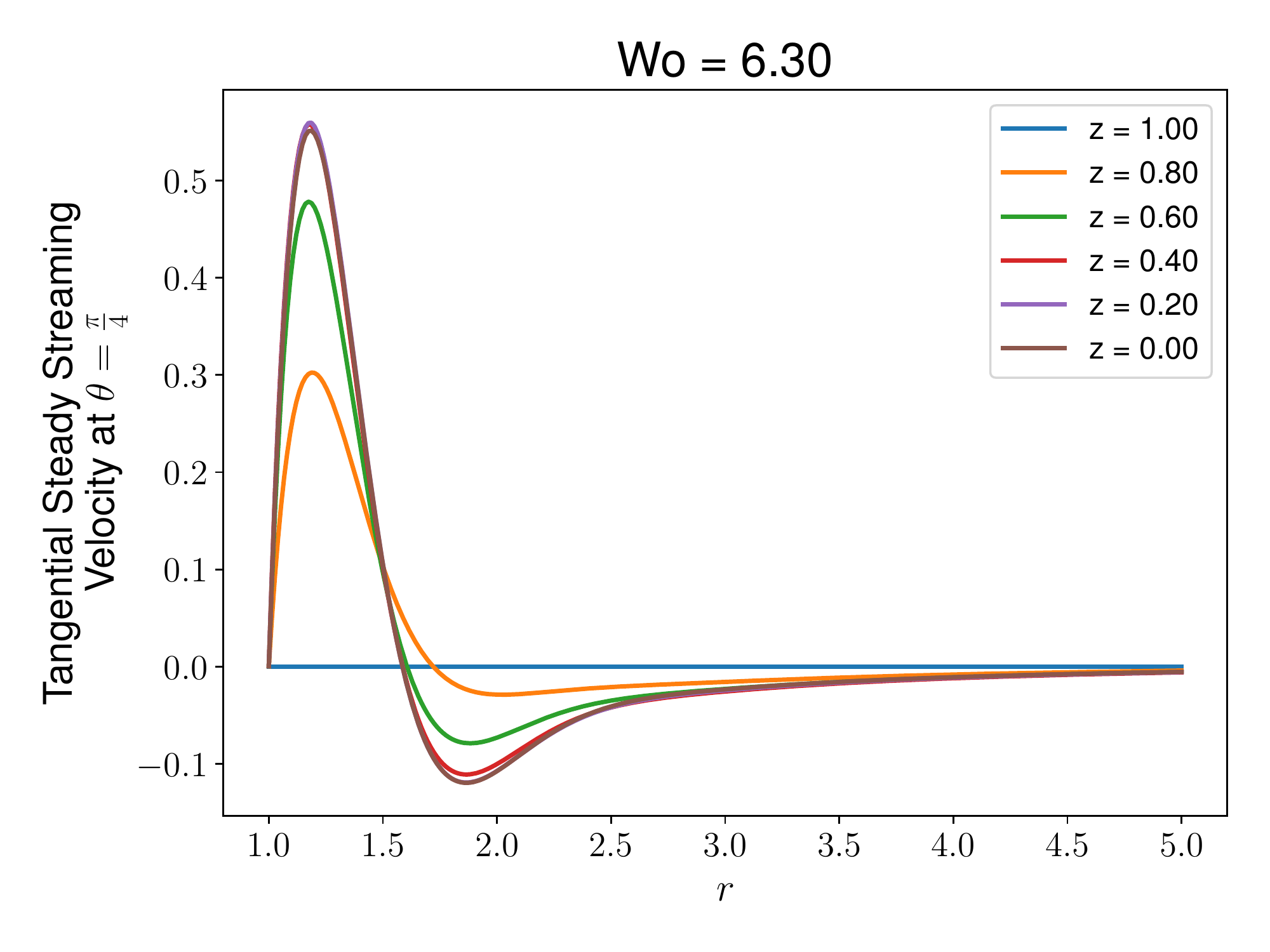}}
    \subfloat[$\Wo \approx 6.3$]{\label{fig:tangVel_znorm}
    \includegraphics[width = 0.24\textwidth,trim = {0.5cm 0.5cm 0.5cm 1.4cm},clip=true]{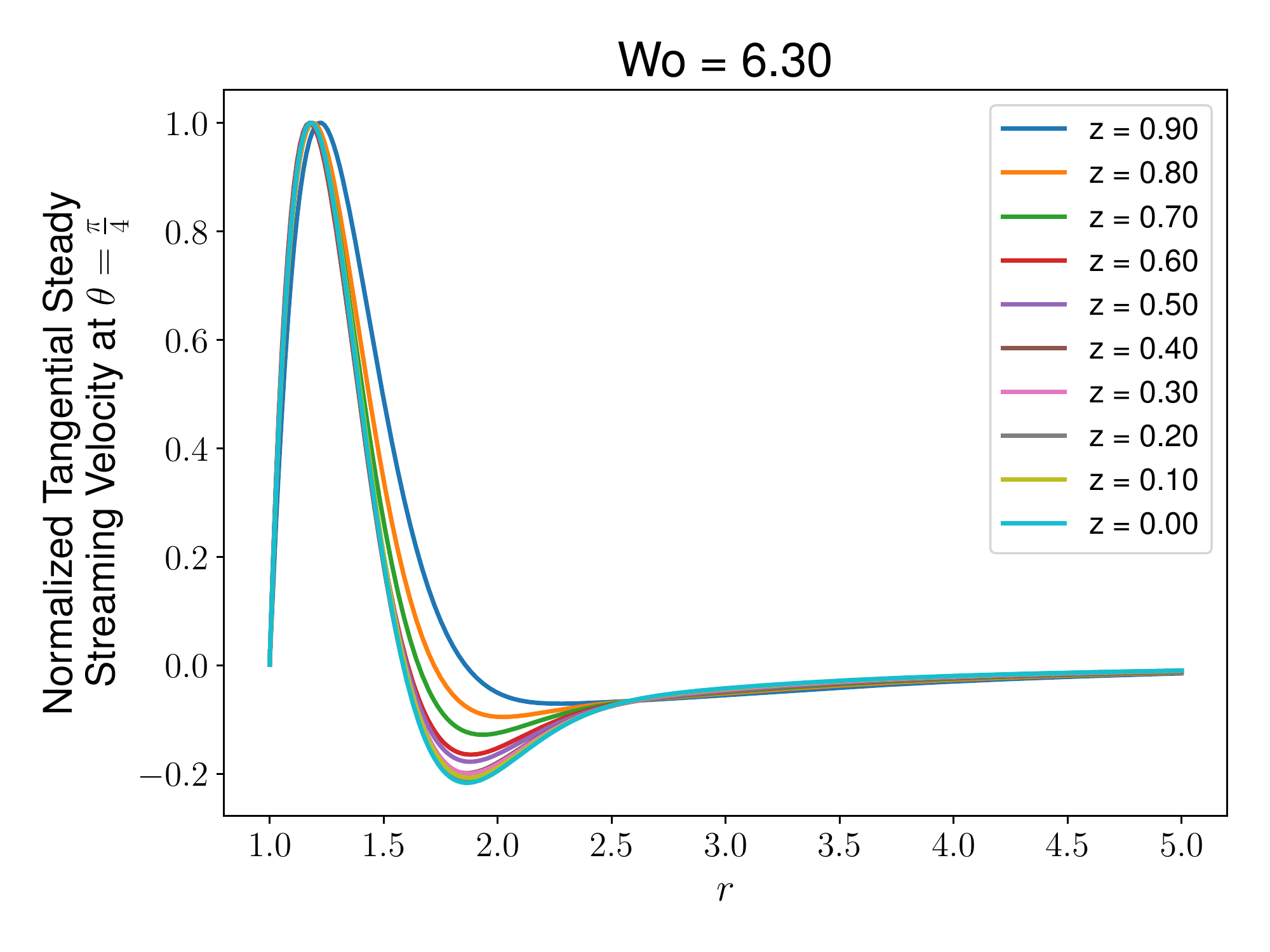}}
\caption{Tangential steady streaming velocity in the azimuthal direction at $\theta = \frac{\pi}{4}$ for $\Wo \approx 2.6,\ 4.8,\ 6.3$.  In plot (d), each curve has been normalized such that the max velocity is 1, for this reason the velocity at the wall $z = \frac{h}{2a}=1$ is omitted.}
\label{fig:TangVel_z}
\end{figure}

We now consider how the tangential steady streaming velocity depends on $z$. In \cref{fig:TangVel_z_Wo2_6,fig:TangVel_z_Wo4_8,fig:TangVel_z_Wo6_3}, we have plotted the tangential steady streaming velocity for six evenly spaced heights between the wall and the middle of the channel for $\Wo \approx 2.6,\ 4.8,\ 6.3$. Again, these plots represent the velocity at which the flow orthogonally crosses the $\theta = \frac{\pi}{4}$ radial line. We note that we have zero velocity at $z=1$ for each $\Wo$ due to the post-processing discussed in \cref{sec:slip}. Further, as the frequency is increased the velocities approach their maximum more quickly at larger values of $z$. This is in agreement with \cref{fig:slipz_vsn2} where we observe the curve flattens in the middle of the channel as $\Wo$ is increased. In \cref{fig:tangVel_znorm}, we have also plotted the tangential steady streaming velocity, now for ten evenly spaced heights from the wall (but not including the wall) to the middle of the channel for $\Wo \approx 6.3$, but in this case, we have normalized all functions such that the max is 1. The purpose of this is to address if the velocity $\u(x,y,z,t)$ can be written as a product $\vect{v}(x,y,t) f(z)$. This is a common technique when considering flows in thin gaps since integrating across the gap leaves only $\vect{v}(x,y,t)$ while retaining some confinement effects. However, implicit in this assumption is that $\u(x,y,z_1,t)$ and $\u(x,y,z_2,t)$ are scalar multiples of one another. However, \cref{fig:tangVel_znorm} shows that evaluating at different $z$ values does not simply return functions that are scalar multiples. Therefore, we conclude that modeling the $z$ dependence of the fluid as a product is not a good assumption for the steady streaming solution in a thin gap.

\begin{figure}[tbhp]
\centering
	\includegraphics[width = 0.5\textwidth,trim = {0cm 0cm 0cm 0cm},clip=true]{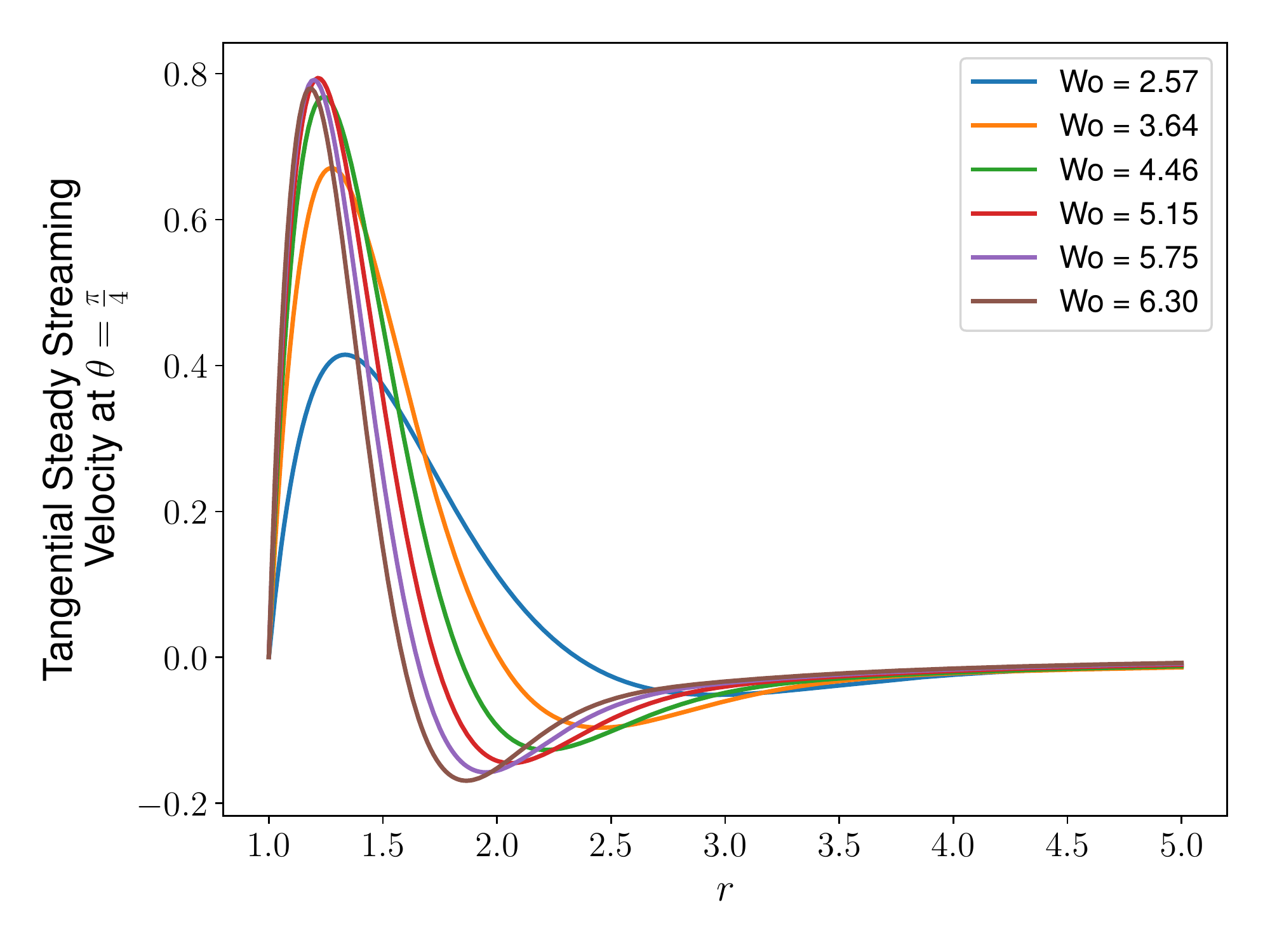}
\caption{Steady streaming velocity in the azimuthal direction at $\theta = \frac{\pi}{4}$ measured at the channel midheight $z = 0$ for $\Wo \approx 2.6,\ 3.6,\ 4.5,\ 5.2,\ 5.8,\ 6.3$}
\label{fig:tangVelOmega}
\end{figure}

We conclude this portion of the results by further analyzing how the steady streaming velocity depends on the frequency, $\omega$. In \cref{fig:tangVelOmega} we have again plotted the tangential steady streaming velocity evaluated at the midheight of the channel for $\Wo \approx 2.6,\ 3.6,\ 4.5,\ 5.2,\ 5.8,\ 6.3$. We note that the maximum velocity grows with increasing $\Wo$ until $\Wo \approx 5.15$ and then decreases slightly, which is consistent with the $\frac{\ell}{a} = 50$ curve in \cref{fig:MaxVelDomain}, More interestingly, we see that the vortex center moves towards the cylinder when the $\Wo$ increases. Here, the vortex center is easily interpreted from \cref{fig:tangVelOmega} as the zero of the tangential steady streaming velocity. Vishwanathan and Juarez \cite{vishwanathan2019steady} experimentally showed that a vortex center in a Newtonian fluid monotonically decreases towards the cylinder as frequency is increased.

\subsection{Higher Order Solution}

\begin{figure}[tbhp]
\centering
	\subfloat[$\Wo \approx 2.6$]{\label{fig:SScorr_2_6}
	\includegraphics[width = 0.32\textwidth,trim = {4cm 0.75cm .25cm 1.5cm},clip=true]{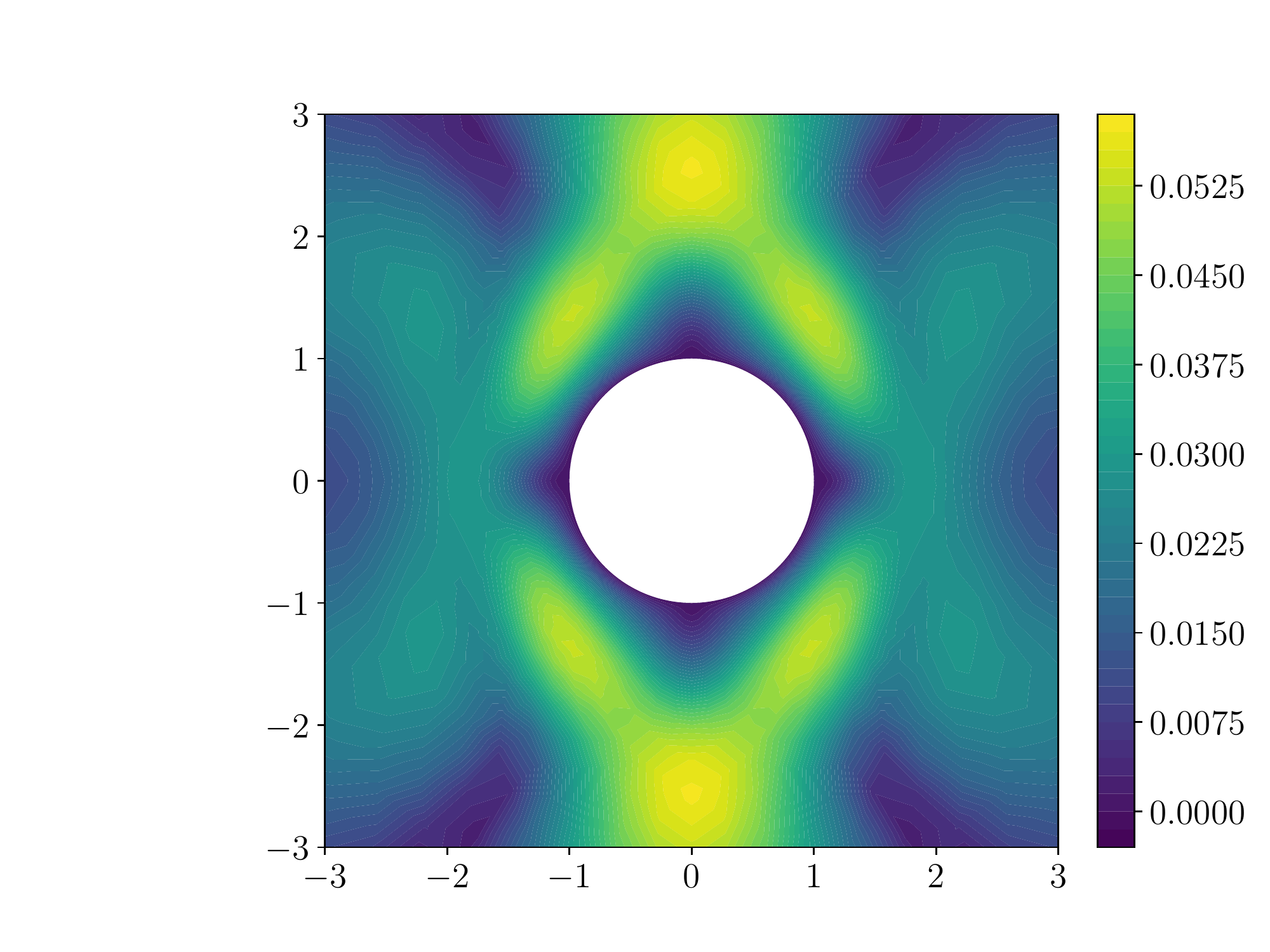}}
	\subfloat[$\Wo \approx 4.8$]{\label{fig:SScorr_4_8}
	\includegraphics[width = 0.32\textwidth,trim = {4cm 0.75cm 0.25cm 1.5cm},clip=true]{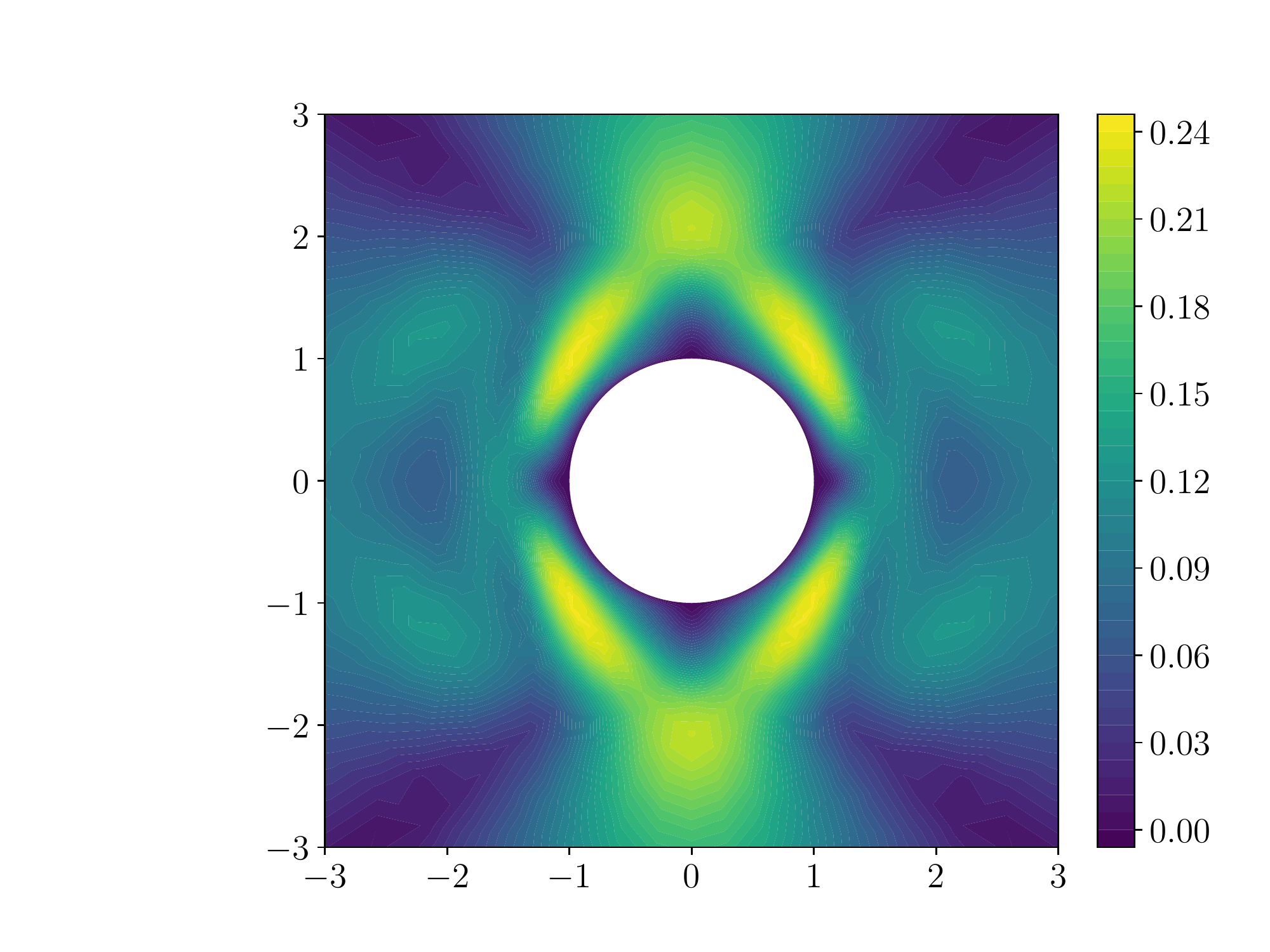}}
	\subfloat[$\Wo \approx 6.3$]{\label{fig:SScorr_6_3}
	\includegraphics[width = 0.32\textwidth,trim = {4cm 0.75cm 0.25cm 1.5cm},clip=true]{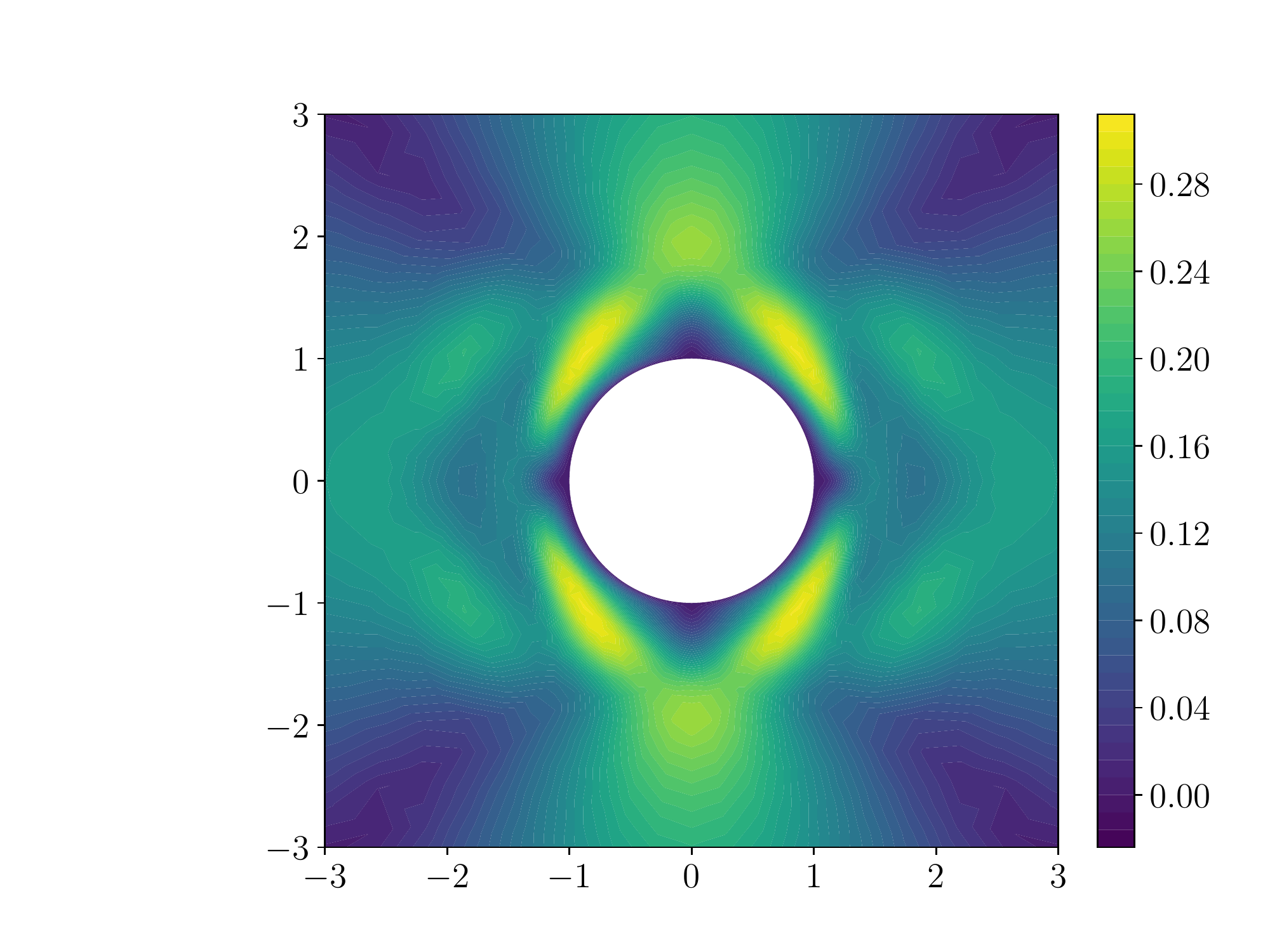}}
\caption{The velocity magnitude of the next order steady streaming solution, $\psi_{0}^{(3)}$, plotted for $\Wo \approx 2.6,\ 4.8,\ 6.3$. We note the areas of high velocity directly above and below (in the $y$-direction) the cylinder such that the solutions no long appear symmetric by a rotation of $\frac{\pi}{2}$. }
\label{fig:SScorr}
\end{figure}

\begin{figure}[tbhp]
\centering
	\subfloat[$\Wo \approx 2.6$]{\label{fig:SScomp_2_6}
	\includegraphics[width = 0.32\textwidth,trim = {4cm 0.75cm 0.50cm 1.5cm},clip=true]{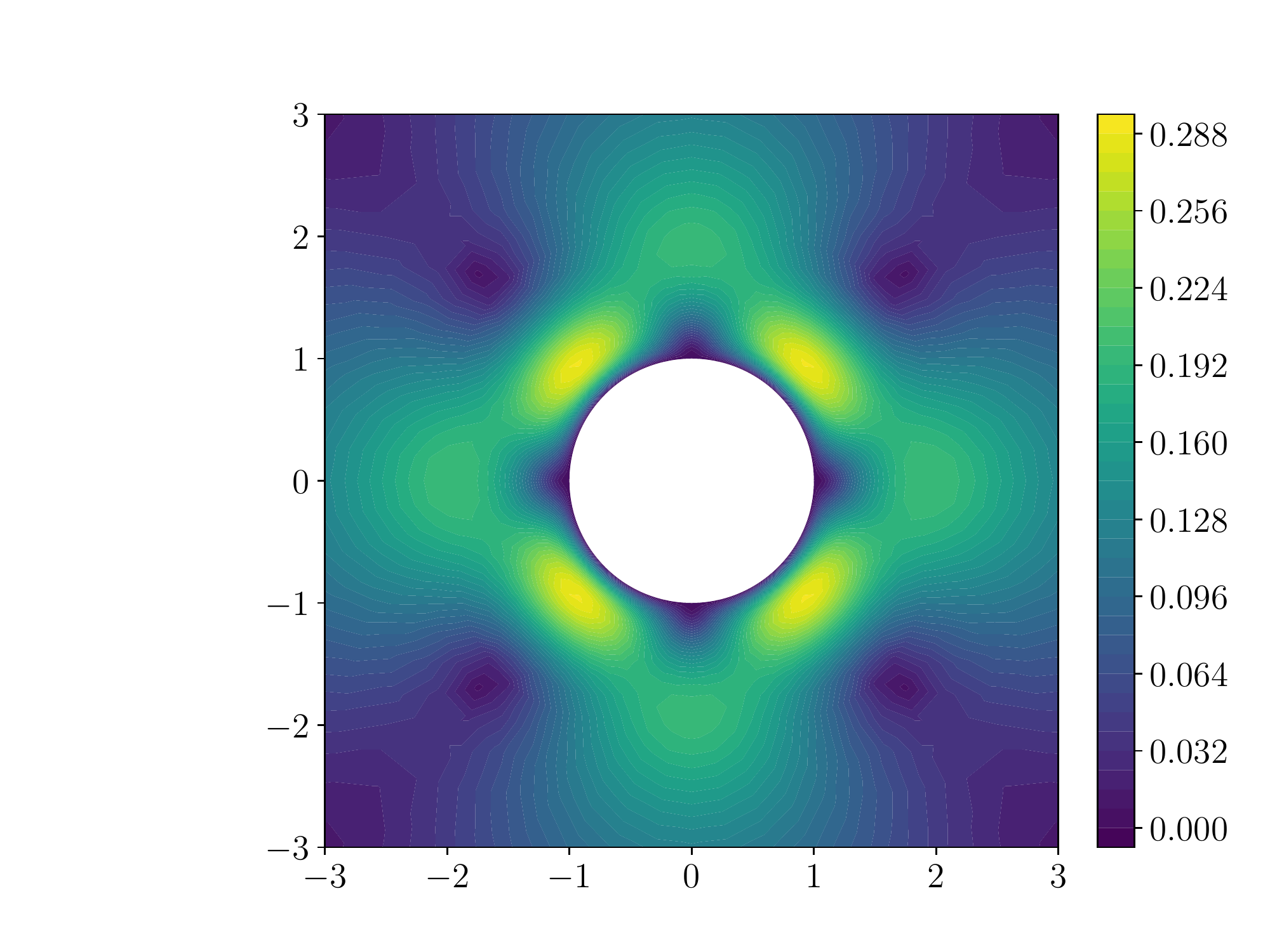}}
	\subfloat[$\Wo \approx 4.8$]{\label{fig:SScomp_4_8}
	\includegraphics[width = 0.32\textwidth,trim = {4cm 0.75cm 0.50cm 1.5cm},clip=true]{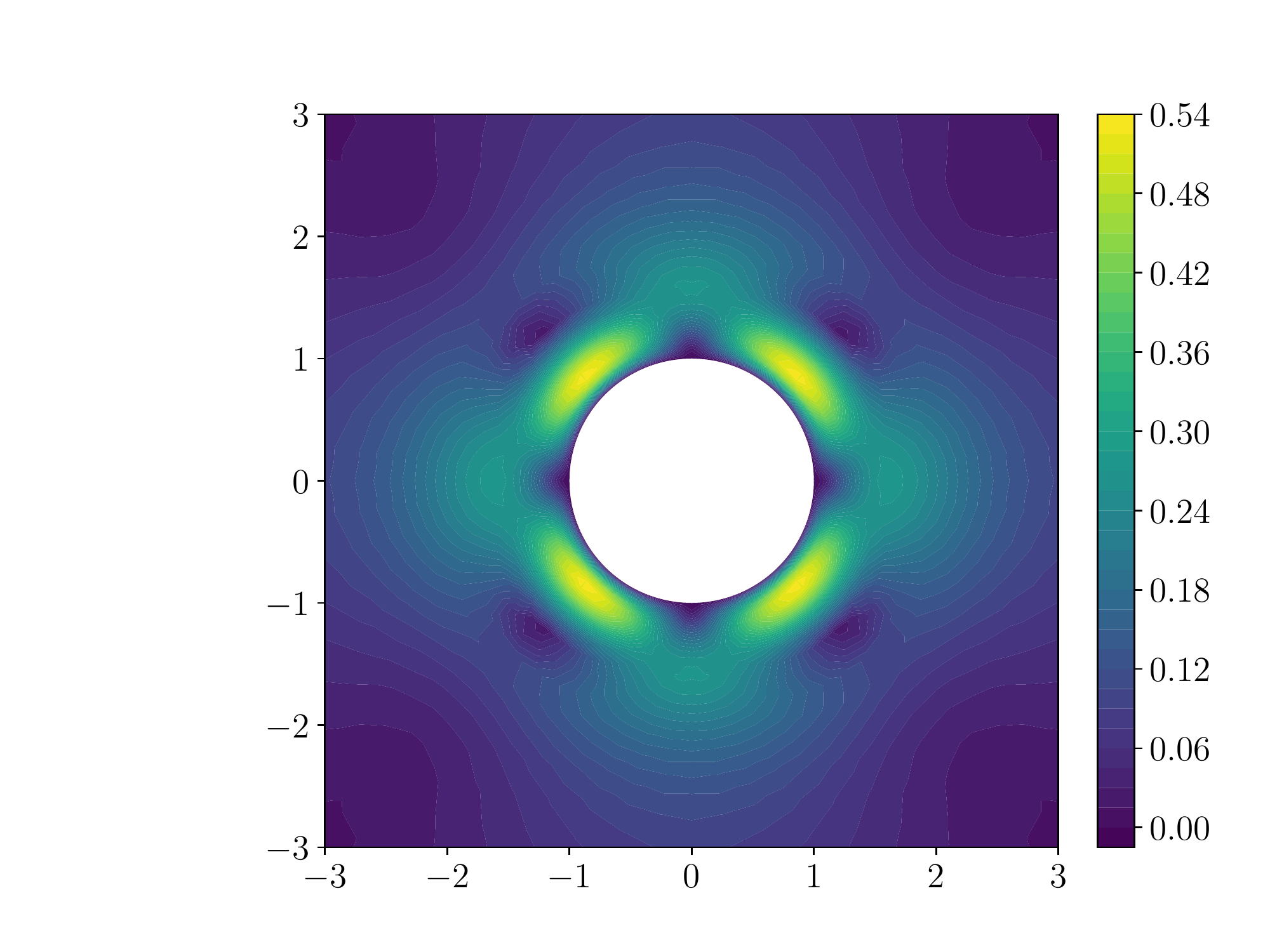}}
	\subfloat[$\Wo \approx 6.3$]{\label{fig:SScomp_6_3}
	\includegraphics[width = 0.32\textwidth,trim = {4cm 0.75cm 0.50cm 1.5cm},clip=true]{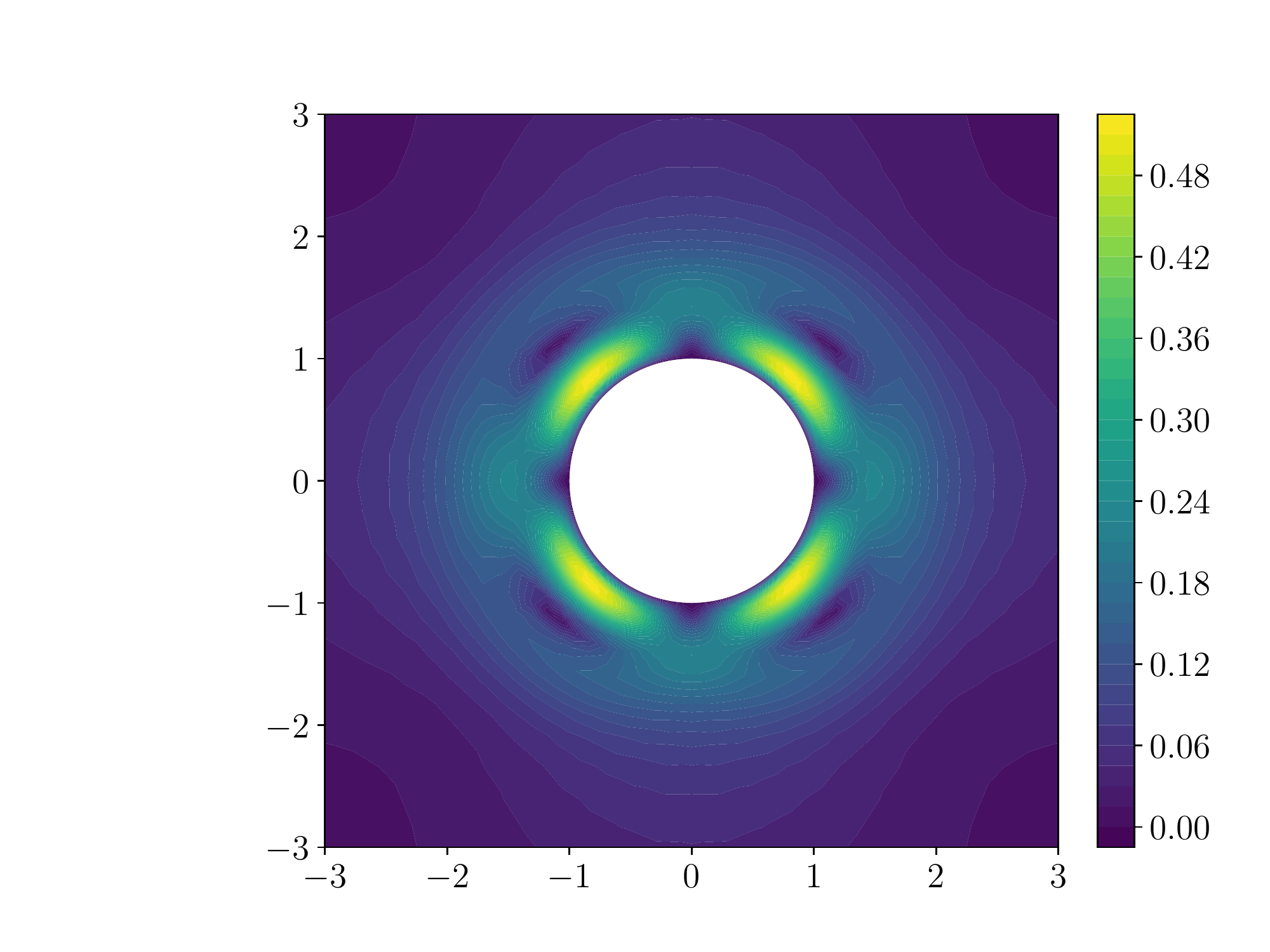}}
\caption{The velocity magnitude of the corrected steady streaming solution, $\psi_{0}^{(1)} + \ep^2\psi_{0}^{(3)}$, plotted for $\Wo \approx 2.6,\ 4.8,\ 6.3$ and $\ep = 0.4$.}
\label{fig:SScomp}
\end{figure}

 The correction term, $\psi_0^{(3)}$, in the asymptotic expansion for the steady streaming solution is plotted in \cref{fig:SScorr}. To compute $\psi_0^{(3)}$ we first must also find $\psi^{(1)}_2$ and $\psi^{(2)}_1$ from \cref{eqn:psi_12,eqn:psi_21}, respectively, and then solve \cref{eqn:psi_30}. There are errors on the rectangular portion of the domain that are small for $\psi_0^{(1)}$, but compound and become significant for $\psi_0^{(3)}$ when using degree 4 polynomials. Therefore, for the higher-order solutions illustrated here we used degree 5 polynomials in our finite element scheme and in this case these errors on the rectangular boundary were inconsequential for $\psi_0^{(3)}$. We note that the apparent symmetry by a rotation of $\frac{\pi}{2}$ is not present in this higher-order solution. This is particularly interesting following the discussion on the symmetry's dependence on the domain. Here, the domain is such that $\frac{L}{a} = 200$ and $\frac{\ell}{a} = 50$, suggesting this symmetry breaking could be a result of the asymptotic expansion and not the shape of the domain whenever the walls are sufficiently far away. For $\Wo \approx 4.8,6.3$, \cref{fig:SScomp_4_8,fig:SScomp_6_3}, the maximum velocities occur on the $y=x$ and $y=-x$ lines similar to $\psi_0^{(1)}$ in \cref{fig:AllQuad,fig:WallTestFixedRatio}, but the solutions are still not symmetric by a rotation of $\frac{\pi}{2}$.

 In \cref{fig:SScomp}, we have plotted the ``corrected" steady streaming solution, $\psi_{0}^{(1)} + \ep^2\psi_{0}^{(3)}$, where we consider $\ep = 0.4$. We note that 0.4 is certainly not {\it much} smaller than one, but it is within the range considered experimentally \cite{vishwanathan2019steady}. We chose an $\epsilon$ where some noticeable change could be observed in the steady streaming solution and was still within the experimental range. For low frequencies, the solutions still appear to be symmetric by a rotation of $\frac{\pi}{2}$, but this is no longer true when $\Wo = 6.3$ as can be seen in \cref{fig:SScomp_6_3}. In \cref{fig:SScomp_6_3} the centers of the vortices and maximum velocities still seem to be rotationally symmetric, but looking closely at the background flow beyond the centers of the vortices it is clear that the velocity magnitude is elongated in the $x$-direction.
 
\section{Discussion} \label{sec:discussion}

In this paper, we have presented the first steady streaming model that includes three-dimensional effects within the classical asymptotic-Fourier steady streaming framework. In this way, we have developed a model that can account for confinement effects and still be easily interpretable with scalar functions. Furthermore, the numerical approximations can be solved quickly, and after solving for the coefficients and reconstructing the $z$-Fourier series, we can easily visualize the flow anywhere within the domain. On the topic of the numerical approximations, to the best of our knowledge, this is the first model that uses a finite element method to find the steady streaming flow within the Fourier series asymptotic expansion structure. The benefit of the finite element method is that our model can easily generalize to other domains. For example, after determining the boundary conditions, we can easily consider the steady streaming flow where the outer boundary is also a cylinder. We could also consider steady streaming induced by oscillatory flow past objects other than cylinders. 

We employed our model to analyze how the steady streaming flow depends on the shape of the domain, the frequency, and the $z$-location within the channel. In particular, we were able to show that as long as the walls are sufficiently far away, we expect the steady streaming flow to retain symmetry by a rotation of $\frac{\pi}{2}$. Our model agrees with previous experiments \cite{vishwanathan2019steady} that the center of the steady streaming vortices moves towards the cylinder as frequency increases. Evaluating the solutions in the $z$-direction, we determined that the velocity is not separable as a function of $z$ and a function of $x$, $y$, and $t$. 

From a modeling and numerical perspective, there is an obvious area to consider for improvements. One criticism is the fact that without post-processing our solutions, we have slip on the $z = \pm\frac{h}{2a}$ walls. We think there is a physically and mathematically interesting optimization problem for the $\beta_k$ in our post-processing scheme. Currently, we take the $\beta_k$ to be constant but want to investigate further to find a set of $\beta_k$ that reduces slip on $z =\pm\frac{h}{2a}$ walls without significantly altering the overall solution. Alternatively, this could be addressed through the choice of the basis. Choosing a basis that strongly enforces no-slip on $z =\pm\frac{h}{2a}$ would be ideal. However, for each basis of this sort that we tried the result was that all the equations for the $z$-Fourier series were coupled together at each asymptotic order creating a significantly less tractable method computationally.

Steady streaming for non-Newtonian fluids has been studied both experimentally \cite{chang1974flow,chang1979secondary,vishwanathan2019steady,vlassopoulos1993characterization} and theoretically \cite{bohme1992steady,chang1977boundary,frater1964secondary,frater1967acoustic,james1977elastico,panda1979harmonically,chang1979secondary,rauthan1969secondary} and has also been utilized for scientific applications \cite{lieu2012hydrodynamic,lutz2006characterizing,thameem2017fast,wang2011size}. While there are several theoretical studies completed for non-Newtonian steady streaming this has all been done with a purely two-dimensional model, same is it was in the Newtonian case. To this point, we are interested in extending our quasi-three-dimensional steady streaming model to non-Newtonian fluids. We are particularly excited at the prospect of theoretically analyzing steady streaming in shear thinning fluids for a wide range of frequencies. 

\section*{Acknowledgements} We thank Akil Narayan for helpful conversations on the numerical approach to the biharmonic equations. We also thank Gabriel Juarez and Giridar Vishwanathan for originally suggesting this topic for research.
\appendix
\section{Separating \texorpdfstring{$\widehat{\bm{F}}_k\bm{(y)}$}{TEXT} into real and imaginary parts.} \label{sec:AppRealImag}

To separate the $F_k(y)$ coefficients into real and imaginary parts we note that the $\lambda_1$ and $\lambda_2$ constants are the only complex terms in \cref{eqn:BCendsFourZ}. Therefore, we define 
$$
    r_k (y) = \frac{\sinh(\sqrt{\lambda_2}y)}{\sqrt{\lambda_2}\cosh\left(\sqrt{\lambda_2}\frac{\ell}{2a}\right)}, \quad 
    s_k = \frac{\sqrt{\lambda_1}\tanh\left(\sqrt{\lambda_1}\frac{h}{2a}\right)}{4a^2\pi^2k^2+h^2\lambda_1},
$$
such that all the remaining terms in \cref{eqn:BCendsFourZ} are real. After splitting $r_k(y)$ and $s_k$ into real imaginary parts the real and imaginary parts of $\widehat{F}_k(y)$ follows. 

We first define $$r_\eta = \left(\left(\frac{a\pi(2m+1)}{\eta}\right)^4 + \Wo^4 \right)^{\frac{1}{4}}, \quad \theta_\eta = \tan^{-1} \left(\left( \frac{\eta \Wo }{a\pi(2m+1)}\right)^2\right),$$ where $\eta$ will be $h$ or $\ell$. Then, using \cite{Mathematica} $r_k(y)$ and $s_k$ separates into real and imaginary parts as 
\begin{align*}
    \Re{r_k(y)} &= \frac{2}{D^2_k}\left(C_3(y) \cos\left( \frac{\ell}{2a} r_h \sin\left(\frac{\theta_h}{2} \right) \right) \cosh\left( \frac{\ell}{2a} r_h \cos\left(\frac{\theta_h}{2} \right) \right) \right. \\
    & \quad \quad \quad \quad \quad \left. + C_4(y) \sin\left( \frac{\ell}{2a} r_h \sin\left(\frac{\theta_h}{2} \right) \right) \sinh\left( \frac{\ell}{2a} r_h \cos\left(\frac{\theta_h}{2} \right) \right) \right) , \\
    \Im{r_k(y)} &= \frac{2}{D^2_k} \left(C_4(y) \cos\left( \frac{\ell}{2a} r_h \sin\left(\frac{\theta_h}{2} \right) \right) \cosh\left( \frac{\ell}{2a} r_h \cos\left(\frac{\theta_h}{2} \right) \right) \right. \\
    & \quad \quad \quad \quad \quad \left. - C_43y) \sin\left( \frac{\ell}{2a} r_h \sin\left(\frac{\theta_h}{2} \right) \right) \sinh\left( \frac{\ell}{2a} r_h \cos\left(\frac{\theta_h}{2} \right) \right) \right), \\
    \Re{s_k} &= \frac{\ell^2 r_\ell}{D^1_k} \left(C^1_k\sin \left(\frac{h}{a} r_\ell \sin\left(\frac{\theta_\ell}{2}\right) \right) + C^2_k \sinh\left(\frac{h}{a} r_\ell \cos\left(\frac{\theta_\ell}{2}\right) \right) \right), \\
    \Im{s_k} &= \frac{\ell^2 r_\ell}{D^1_k} \left(C^2_k \sin\left(\frac{h}{a} r_\ell \sin\left(\frac{\theta_\ell}{2}\right) \right) - C^1_k \sinh\left(\frac{h}{a} r_\ell \cos\left(\frac{\theta_\ell}{2}\right) \right) \right),
\end{align*}
where 
\begin{align*}
    C^1_k &= \left( h \ell \Wo \right)^2 \cos\left(\frac{\theta_\ell}{2}\right) - \left(a \pi \right)^2 \left( \left(2k\ell\right)^2 + \left(h\left(2m+1\right)\right)^2 \right) \sin \left(\frac{\theta_\ell}{2} \right), \\
    C^2_k &= \left(a \pi \right)^2 \left( \left(2k\ell\right)^2 + \left(h\left(2m+1\right)\right)^2 \right) \cos \left(\frac{\theta_\ell}{2} \right) + \left( h \ell \Wo \right)^2 \sin\left(\frac{\theta_\ell}{2}\right), \\
    C^3_k(y) &= \sin\left(\frac{\theta_h}{2} \right)\sin \left(r_h y \sin \left(\frac{\theta_h}{2} \right)\right) \cosh\left(r_h y \cos \left(\frac{\theta_h}{2} \right) \right) \\ 
    & \quad \quad + \cos\left(\frac{\theta_h}{2} \right)\cos \left(r_h y \sin \left(\frac{\theta_h}{2} \right)\right) \sinh\left(r_h y \cos \left(\frac{\theta_h}{2} \right) \right), \\
    C^4_k(y) &= \cos\left(\frac{\theta_h}{2} \right)\sin \left(r_h y \sin \left(\frac{\theta_h}{2} \right)\right) \cosh\left(r_h y \cos \left(\frac{\theta_h}{2} \right) \right) \\ 
    & \quad \quad - \sin\left(\frac{\theta_h}{2} \right)\cos \left(r_h y \sin \left(\frac{\theta_h}{2} \right)\right) \sinh\left(r_h y \cos \left(\frac{\theta_h}{2} \right) \right), \\
    D^1_k &= \bigg( (a \pi)^4 \left( \left(2k\ell\right)^2 + \left(h\left(2m+1\right)\right)^2 \right)^2  \\ & \quad \quad + (h \ell \Wo)^4 \bigg) \left( \cos\left(\frac{h}{a} r_\ell \sin \left( \frac{\theta_\ell}{2}\right)\right) + \cosh\left(\frac{h}{a} r_\ell \cos \left( \frac{\theta_\ell}{2}\right)\right) \right), \\
    D^2_k &= r_h \left( \cos \left( \frac{\ell}{a} r_h \sin \left( \frac{\theta_h}{2} \right) \right) + \cosh\left( \frac{\ell}{a} r_h \cos \left( \frac{\theta_h}{2} \right)\right)\right). \\
\end{align*}

Executing the hyperbolic functions led to overflow errors, therefore to implement these functions we expand the hyperbolic functions via there exponential definitions. Then, we factor out the exponentially large terms from each and divide out these problematic terms when simplifying the fractions. 

\printbibliography

\end{document}